\documentclass[12pt,a4paper,twoside,ngerman,pagesize=auto,headinclude=true,footinclude=true, DIV=12,BCOR=0cm]{scrartcl}
\usepackage{graphicx}
\usepackage{epstopdf, epsfig}
\usepackage{gensymb}
\usepackage{subcaption}
\usepackage[round,authoryear]{natbib}
\usepackage{amssymb}
\usepackage{amsbsy}
\usepackage{scrlayer-scrpage}
\usepackage[left=2.0cm,right=2.0cm, top=2.5cm,bottom=3.0cm]{geometry}
\usepackage{afterpage}

\title{Coherent structures in the wake of a SAE squareback vehicle model}
\author{Benjamin Bock \footnote{Email address for correspondence: Benjamin.Bock@tplus-engineering.de} \\ 
Tplus Engineering GmbH, \\ Steinbeisstra\ss{}e 25, \\ 70771 Leinfelden-Echterdingen, Germany
}
\date{}

\clearpairofpagestyles

\lehead{\thepage}
\rohead{\thepage}
\cehead{B.Bock}
\cohead{Coherent Structures in a Vehicle Wake}
\begin{document}

\maketitle

%\begin{center}
%\textbf{Benjamin Bock}
%
%Tplus Engineering GmbH, Steinbeisstra\ss{}e 25, 70771 Leinfelden-Echterdingen, Germany
%\end{center}

\begin{abstract}
The wake of a SAE squareback vehicle model is studied both experimentally and numerically for a Reynolds-number of $ Re_h = 1.0 \hspace{5pt} 10^6 $.The investigation focuses on the coherent structures of the intermediate to largest length and time scales. Flow field as well as base pressure fields are observed for the understanding of the relation between the signals of these quantities. Generalizations and differentiations are made by comparison with the documented behavior of Ahmed or similar vehicle models or three-dimensional bluff bodies. In comparison the vortex shedding acts similar but is restricted to the upper half of the wake of the SAE vehicle model. Due to the localization and phase behavior of the vortex shedding the connection between the base pressure signals and the flow field is weak. However, the pressure signals may be a viable feedback sensor under certain conditions, for example in flow control applications. A flapping of the near wake is identified for the fluctuations of the low frequency time scales. \end{abstract}

\vspace{12pt}


\section{Introduction}
In recent decades, the manipulation of the flow in the wake of bluff bodies has been a topic of great interest in numerous investigations. Many examples can be found in the shape optimization of vehicles \citep[see comprehensive work of][]{hucho1993aerodynamics} and buildings \citep[see comprehensive work of][]{xie2014aerodynamic}, as well as in approaches to control these and similar basic bluff body flows in a passive or active manner \citep[][]{Choi2008control, kim2008passive, Pastoor2008, Wassen2010, Krentel2010, Barros2015, schmidt2015drag, wieser2015manipulation, littlewood2012aerodynamic}. The main objective of these efforts is to control the forces of the bluff body in a desired direction or way. This means in particular to reduce drag, to increase or decrease lift or to control the distribution or the dynamics of forces. The latter is relevant, for example, in the vibration of buildings \citep[e.g.][]{hayashida1990aerodynamic, xu1992control} and in the steering of a vehicle \citep[][]{stoll2015unsteady}. For the manipulation of such flows around bluff bodies the understanding of the dynamics and the motion of the flow in the wake is essential.  

In various ways the literature \citep[see][]{Pastoor2008,Barros2015,Bock2016} demonstrates that fluctuations in the wake flow caused by vortex motions or coherent structures are sources of losses which contribute to aerodynamic drag. Hence the reduction or the prevention of coherent structures will lead to a reduction of aerodynamic drag. However, the dynamics of coherent structures may partially depend on the form of the bluff body. This study focuses on the geometry of a squareback vehicle and the detailed observation of occuring coherent motions of fluctuations in the wake.  

Simliar to two-dimensional flows, mainly $3$ different forms of coherent structures exist in the wake of a squareback vehicle based on the time and length scales. Firstly, vortices roll up emanating from the Kelvin-Helmholtz (KH) instability in the shear layer. Secondly, comparatively larger vortices shed from the dead water region. The vortex shedding occurs in time scales of one order of magnitude above the formation of KH vortices. Thirdly, within time scales of at least one order of magnitude above vortex shedding, the entire dead water region experiences deflections. The following paragraphs summarize the documented observations of these coherent structures on vehicles and vehicle models so far. 

Kelvin Helmholtz (KH) vortices emerge in the shear layer of almost every wake flow. \citet[][]{Barros2015} detects frequencies between $Sr = f \delta_S / u = 0.23$ and $Sr = 0.29$ (related to the local shear layer thickness $ \delta_S $ and the velocity $ u = 0.5 u_{\infty} $, with the freestream velocity $ u_{\infty} $) of an Ahmed vehicle model in a  Reynolds-numbers range of $ Re_h = u_{\infty} h / \nu = 2 \hspace{5pt} 10^5 $ and $ Re_h = 4  \hspace{5pt} 10^5 $ (with the vehicle height h and the kinematic viscosity $\nu$). The peaks of the KH mode are broadband and of decreasing frequency with increasing distance from the separation edge \citep[][]{Barros2015}. The decrease of frequency with distance is interpreted as a consequence of the vortex pairing. \citet[][]{Grandemange2014c} investigates the shear layers in the wakes of two production cars at a Reynolds-number of $ Re_l = 10^7 $. These investigations show the phenomenon of vortex pairing of shear layer vortices in terms of a linear growth of shear layer in flow direction and proportional decrease of frequency of the dominant part of fluctuations.  

Vortices are permanently shed from the near wake into the far wake. Publications related to squareback vehicle models of \citet[][]{Grandemange2013} at Reynolds-number $ Re_h = 9.2 \hspace{5pt} 10^4 $ and \citet[][]{Barros2015} at Reynolds-number $ Re_h =  3 \hspace{5pt} 10^5 $ detect the frequency of vortex shedding between $ Sr_h = f h / u_{\infty} = 0.127 $ und $ Sr_h = 0.215 $. The structures of vortex shedding of opposing shear layers in the horizontal plane are shifted by half a wavelength \citep[][]{Grandemange2013}. These observations agree with the results of axiymmetric bluff bodies and elliptic flat plates in cross flow \citep[][]{Rigas2014, Grandemange2012, Grandemange2012a, Kiya1999, yang2015low}. In the orthogonal plane to the width of the vehicle the distance of structures grows linear with the distance to the vehicle base. This behavior is attributed to the influence of the ground proximity and the different propagation velocities of upper and lower shear layer \citep[see][]{Grandemange2013}. Hence, \citet[][]{Grandemange2013} and \citet[][]{duell1999experimental} report a meandering, stretched vortex loop with respect to the vortex shedding of squareback vehicle models based on their measurements. The same frequencies of such global modes are detected in the shear layers of 2 production cars by \citet[][]{Grandemange2014c}. 
 
Low frequency fluctations appear as a slight shift of the orientations of the near wake relative to the mean flow direction. Up to now the literature has reported these low frequency fluctuations of the wake only in terms of a symmetry breaking mode through a bi-stability \citep[][]{Grandemange2013, Grandemange2013a, brackston2016stochastic}. In case of a bi-stable behavior the wake is deflected and switches direction sponaneously after long time scales (compared to the typical time scales of the vortex shedding \citep[see][]{Grandemange2013}). The averaged timescale of the switches is approximately $ \Delta t u_{\infty} / h = 1500 $ and scales with the velocity. The geometric proportions of the vehicle like the ratio of height to width h / b and the ground clearance determine whether a bi-stability occurs and the axis of the bi-stability is aligned with the height or the width of the vehicle \citep[][]{Grandemange2013a}. The investigations of \citet[][]{Grandemange2014c} on 2 production cars and those of \citet[][]{Cadot2016} on 4 production cars for a Reynolds-number of $ Re_l = u_{\infty} l / \nu = 10^7 $ (with the vehicle length l) show that bi- stabilities exist under realistic conditions. However, they also show that they may not always occur.

Some approaches of active flow control reveal aspects of the behavior, the structures and local phase relations of detached vortex structures in the wake of squareback vehicles. The response of symmetric and antisymmetric excitation on the separation edges of an Ahmed model with synthetic jets confirms the vortex shedding in a meandering form \citet[][]{Barros2016resonances}. \citet[][]{Rigas2014, rigas2017weakly} observe a comparable response for axialsymmetric bluff bodies. The excitation of the vortex shedding results in a reduction of fluctuating energy for other coherent structures \citep[][]{rigas2017weakly}. This indicates that the dynamics of different coherent structures are affecting each other. 

There are still many geometrical influences where only little is known about the effect on coherent structures and the vortex shedding. Some of these are the front shape (e. g. without separation bubble or rather sharp front shape), a rear diffuser or the moving ground. \citet[][]{Barros2016resonances} mention in their work that further studies for the influence of different geometries (e.g. aspect ratio of the base, ground clearance) are necessary for a better understanding of the vortex shedding. There is also no reported study for the fluctuation motions at long time scales in the case when no bi-stable behavior is detected. However, the occurence of a bi-stable behavior depends on geometrical influences \citep[][]{Grandemange2013a}. Hence the question concerning the influence of geometrical parameters on vortex shedding can be extended to low frequency behavior. 

With respect to flow control of fluctuating motions the understanding of processes in the wake flow and their relation to the behavior of base pressure fluctuations is important. This applies especially to possible different types of behaviors dependent on geometrical parameters. An example of an effective control of coherent structures in two-dimensional flow shows the importance of the understanding of this relation \citep[][]{Pastoor2008}. Recent results on active flow control on three-dimensional bluff bodies \citep[][]{rigas2017weakly} and on an Ahmed squareback model \citep[][]{Barros2016resonances} support this statement. These studies show that the understanding of coherent structures and their phase relation on different positions as well as the direction of motion are essential to create concepts for an effective control. Furthermore, the example of the two-dimensional flow shows particularly that the understanding of the relation between dynamics of coherent structures and the resulting base pressure behavior can be valuable for sensor design and position for an effective feedback loop flow control \citep[][]{Pastoor2008}. 

This study investigates coherent structures in the wake of a SAE squareback vehicle model as an approach to address these open questions. The focus is on the vortex shedding and fluctuating motions at low frequencies. The coherent structures of such a model and its particular geometric characteristics have not been documented until now. The SAE squareback model represents a geometry with a rear diffuser and with an aspect ratio of the base that differs from the geometries that were investigated so far \citep[][]{Grandemange2013, duell1999experimental}. Together with these results the present study gives an impression of the possible impact of these geometrical details on coherent structures.

The present work is structured as follows. Section 2 presents methods of filtering and separating coherent structures as well as the motion of structures from other remaining non-coherent turbulent fluctuation motions. The following section describes the experimental setup and the measurement configuration for data acquisition. Section 4 introduces the numerical model that is used to extend the experimental observation of results. The results of the study are presented in section 5. First the results on base pressure and then on the flow field are shown. These results are further subdivided by different time and length scales. In the discussion in section 6 these results are compared to findings from other investigations and extended to more typical production car shapes provided by additional experimental data. The results of this present study are summarized in section 7. This last section provides an outlook related to the impact on the initial questions in section 1 and concludes with ideas for further research to answer remaining questions. 

\section{Methods}
The investigation of coherent structures in wakes requires a concept for coherent structures as well as the application of methods that allow a distinction of different coherent structures. This section discusses the interpretation of coherent structures in the present investigation. The remaining part of this section is dedicated to the applied methods for an identification and a description of coherent structures. 

\subsection{Coherent Structures}
Coherent structures are related to large scale vortex-like structures. However, the description of coherent structures differs in some details from the definition of vortices and turbulence. Many examples for the utilization of the concept of coherent structures can be found in research related to turbulent flows \citep[][]{Berger1990, Hussain1983, Hussain1986, Lee2004, Rempfer1991, Pastoor2008, Pujals2010, sirovich1987turbulence, fuchs1979large, adrian2000analysis}. The example of the intuitive definition of a vortex by closed, circular pathlines around a center only leads in some special cases to the identification of vortices \citep[][]{Hussain1986, adrian2000analysis}. The weak definition of coherent structures offers many possibilities compared to the definition of a vortex. This serves as a possibility to a simpler access in their observation. To clarify the description of coherent structures for the present study the following list gives a weak definition of the term. This definition follows mainly \citet[][]{Hussain1986} and \citet[][]{ho1984perturbed}. Coherent structures are essentially related to:   

\begin{itemize}

   \item The occurrence of a temporal or spatial coherence, orderliness or structured behavior relative to the temporal or spatial average. 

   \item Vorticity, or vortical structures.

   \item \begin{flushleft} Fractions of large scale or low frequency fluctuations in the turbulence spectra (in comparison to the dissipative scales).\end{flushleft}

   \item A spatial coherence that may appear as an unsteady, periodic but also intermittent process. 

\end{itemize}

The following part of the section presents methods for the description and observation of coherent structures. 

\subsection{Filtering of small scales and frequencies}
Small scale fluctuations of a pressure field are separated from large scale fluctuations by observation of the barycenter (equation \ref{SP}, \citep[see][]{Grandemange2013}) or a dimensionless, spatially averaged surface pressure gradient (equation \ref{Grad}, \citep[see][]{Grandemange2013a}). This results in two values for each quantity ($2$ coordinates or directions of the gradient) for a given pressure distribution. These values characterize the considered domain in terms of large scale motions. The barycenter is determined by the quotient of the integral from the product of the local base pressure and its position, divided by the integral of the base pressure. The dimensionless, spatially averaged surface pressure gradient is computed by the sum of the product of local average of pressure along a constant coordinate $ c_{pz,i} $ or $ c_{py,i} $ with the local coordinate $ y_i $, $ z_i $ relative to the maximal distance of measurement point locations $ \Delta y_{max} $, $ \Delta z_{max} $.

\begin{equation} \label{SP}
	y_p = \frac{\int y c_p ds}{\int c_p ds}; \hspace{20pt} z_p = \frac{\int z c_p ds}{\int c_p ds}; 
\end{equation}
\begin{equation} \label{Grad}
	\frac{\partial c_p}{\partial y} = \frac{\Sigma c_{pz,i} y_i}{\Delta y_{max}}; \hspace{20pt} \frac{\partial c_p}{\partial z} = \frac{\Sigma c_{pz,i} y_i}{\Delta z_{max}}; 
\end{equation}

Periodically occuring processes are considered by power spectral densities (PSD, $ S_{xx}(f)$) in frequency space. Hence, period times T represent frequencies f. The PSD is computed using the autocorrelation function $r_{xx}(\tau)$ (equation \ref{Auto}) of the time row $x(t)$ which folds the time row to different time lags $ \tau $. The PSD (equation \ref{PSD}) is calculated by Fourier transformation of the autocorrelation function $r_{xx}(\tau)$ . The impact of random fluctuations is reduced through averaging over overlapping windows of autocorrelation functions. Spectral leakage through the limited window is avoided by window functions.

\begin{equation} \label{Auto}
	r_{xx}(\tau) = \lim\limits_{T \rightarrow \infty}{\int_{-T}^{T} x(t) x(t+\tau) dt}
\end{equation}  
\begin{equation} \label{PSD}
	S_{xx}(f) = \frac{1}{2 \pi} \int r_{xx}(\tau) e^{-j2 \pi f \tau} d\tau
\end{equation}

To assess peak frequencies the frequency f must be available in a meaningful format. This is the Strouhal number (equation \ref{Sr})

\begin{equation} \label{Sr}
	Sr_h = \frac{f h}{u_{\infty}}
\end{equation}

\subsection{Proper Orthogonal Decomposition}
The Proper Orthogonal Decomposition (POD) is utilized here to identify spatial structures and related time scales as well as for a Low Order Modelling (LOM). The separation of coherent structures is based on the POD method of snapshots \citep[][]{sirovich1987turbulence}. Starting point is the decomposition of fields from a number $ M $ of snapshots $ u(x,t) $ into fluctuating fields $ u'(x,t) $ and an averaged field $ \bar{u}(x) $. The POD decomposes the fluctuating fields in $ M $ spatial modes $ \phi_m (x) $ and temporal coefficients $ a_m(t) $:

\begin{equation} \label{POD}
	u'(x,t) = \sum_{k=1}^{M} a_m(t) \phi_m(x).
\end{equation} 

The spatial modes describe the spatial distribution of the fluctuations and the temporal coefficients reflect their time-dependent amplitude. The modes $ \phi_m(x) $ correspond to the eigenvectors from the Singular Value Decomposition (SVD) of the autocorrelation matrix $ R(x) $.

\begin{equation} \label{AutoCorMatrix}
	R(x) = \overline{u'(x,t) \cdot u'(x',t)}.
\end{equation}

The eigenvalues $ \lambda_k $ of the SVD represent the energy fraction of the turbulent kinetic energy of the fluctuations. In the process of the SVD, the eigenvalues are normalized, so that $ \| \phi_k \| = 1 $ and the modes are sorted by descending size (energy  fraction). The temporal coefficients are determined by the projection of the fluctuations on the modes:

\begin{equation} \label{koeff}
	a_k(t) = \phi_k(x) \times u'(x,t).
\end{equation}

The time coefficients are considered in a spectral decomposition to analyze the time scales of the modes. In addition, the spatial, time-dependent behavior and fluctuations of the modes are investigated. Lower Order Models (LOM) are used here. A LOM can consist of one or more modes with $m = 1$ to $m = M$. Equation \ref{LOM} represents the reconstruction of a field with a LOM:

\begin{equation} \label{LOM}
	u_{LOM}(x,c) = \bar{u}(x) + \sum_{m=1}^{M} c_m \phi_m(\vec{x}).
\end{equation}

The averaged field $ \bar{u}(x) $ is superimposed with the product from the mode $ \phi_m $ and the amplitude $ c_m $ for all modes taken into account. The limitation of the number of modes can be considered as a spatial filter. Since spatially small structures in the turbulent spectra usually correspond to a small energy fraction of the fluctuations, higher mode numbers $m$ tend to represent smaller spatial structures (and higher frequencies). In the simplest case, the amplitude $ c_m $ corresponds to a selected amplitude $ a_m(t) $ at a selected time instant $t$. In more complex LOMs, an appropriate definition must be given.

Mode structures of two modes, which represent the same size of structures shifted by a quarter of a wavelength in the flow direction, are interpreted and modeled as a convective mode pair. In time-resolved data sets, the same time scales occur for these modes in convective mode pairs in the time coefficients. This is used for the LOM to describe a process and define the amplitudes $ c_m $. For the LOM of a convective mode pair, the phase definition based on the time amplitudes $ a_1(t) $ and $ a_2(t) $ as used by \citet[][]{chiekh2013pod} and given by equation \ref{phase} is applied:

\begin{equation} \label{phase}
	\varphi_{1,2}(t) = \arctan{ \frac{\sqrt{2} \lambda_1 a_1(t)}{\sqrt{2 \lambda_2} a_2(t)}}.
\end{equation}

For eigenvalue based, phase-dependent amplitudes, the following applies:

\begin{equation} \label{PhAmplituden}
	c_1(t) = \sqrt{2 \lambda_1} \sin{\varphi_{1,2}(t)}, \hspace{20pt} c_2(t) = \sqrt{2 \lambda_2} \sin{\varphi_{1,2}(t)}.
\end{equation}

Due to disordered, turbulent and intermittent fluctuating motions, different amplitudes $ c_1(t) $, $ c_2(t) $  exist for the same phase $ \varphi(t) $ at different times $t$. A phase-averaged amplitude $ c_1 $, $ c_2 $ can be calculated, which can be used as LOM to describe a phase-averaged process. In some cases, the higher order modes $m=3$ and $m=4$ are also included in the phase averaging based on the phase definition of equation \ref{phase}: 
\begin{equation} \label{PhAmplituden34}
	c_3(t) = \sqrt{2 \lambda_3} \sin{\varphi_{1,2}(t)}, \hspace{20pt} c_4(t) = \sqrt{2 \lambda_4} \sin{\varphi_{1,2}(t)}.
\end{equation}

\section{Experimental setup on the wake measurements of a SAE model}
The coherent structures in the wake of an SAE model were experimentally investigated in a wind tunnel. The geometry of the SAE squareback model is described by \citet[][]{lindener1999aerodynamic}. The wind tunnel design, measurement technology and data recording used are described below.

\subsection{Geometry and wind tunnel setup}

\begin{figure}[h]
%  \centerline{\includegraphics[width=11.54cm]{Bilder/ExperimentellerAufbauSAEModell.png}}% Images in 100% size
  \centerline{\includegraphics[width=11.54cm]{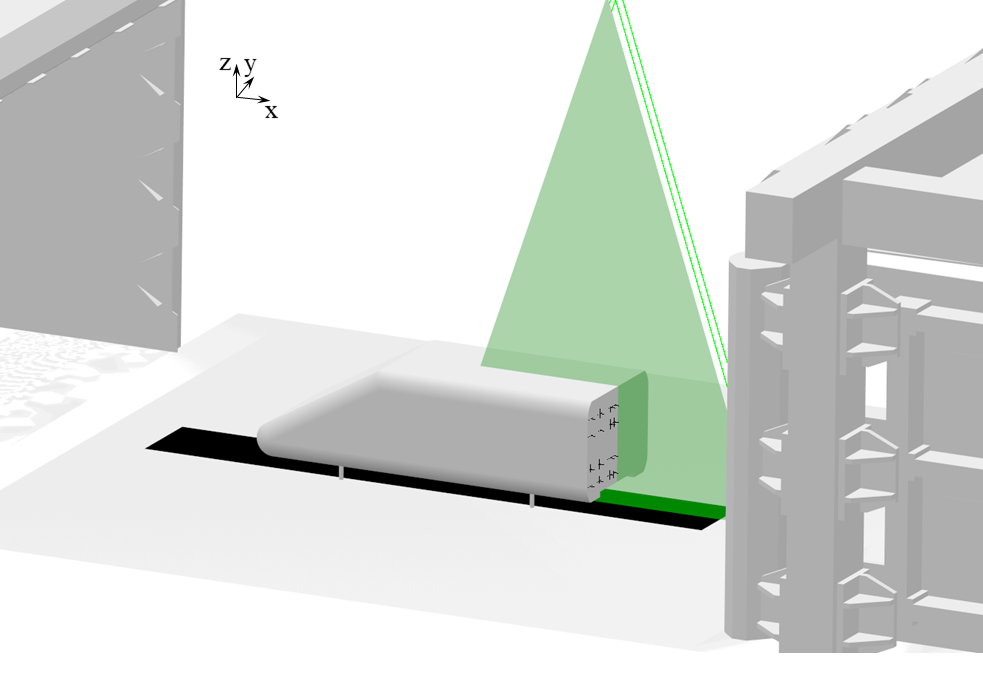}}% Images in 100% size  
%  \centerline{\includegraphics[width=14cm]{Bilder/ExperimentellerAufbauSAEModell.png}}% Images in 100% size
  \caption{Experimental setup of the SAE squareback model in the wind tunnel}
\label{ExpSetup}
\end{figure}

The experimental investigations on a $1:4$ SAE squareback model were carried out in the model wind tunnel of the Institute for Internal Combustion Engines and Automotive Engineering (in German Institut f\"ur Verbrennungsmotoren und Kraftfahrwesen: IVK) at the University of Stuttgart. The wind tunnel is equipped with a system for simulating the moving ground, which includes a five belt system and a boundary layer pre-suction system. A block profile of the flow is generated with the boundary layer pre-suction and the moving floor. For the velocity ranges between $ u \in [ 30;80 ] \hspace{5pt} m \hspace{5pt} s^{-1} $ the turbulence intensity is $ Tu < 0.3 $. Further details on the wind tunnel are described in \citet[][]{wiedemann2003new}. The measurements were carried out for the velocities $ u =\{30, 40, 50, 60, 70\}  \hspace{5pt} m \hspace{5pt} s^{-1} $ and correspond to the Reynolds number $ Re_h = \{0.6, 0.8, 1.0, 1.2, 1.4\} \hspace{5pt} 10^6 $. In this publication only the results at $ Re_h = 1.0 \hspace{5pt} 10^6 $ are described. However, the same behavior was observed for all Reynolds numbers. The arrangement of the model and the measuring technique is shown schematically in figure \ref{ExpSetup}. The picture shows the nozzle outlet on the left and the collector on the right with the SAE model in between. The coordinate system is defined in such a way that the $x$-axis runs parallel to the flow, the and the $z$-axis runs parallel to the height direction with the origin in the middle of the base surface.

%\afterpage{\clearpage}

The reference velocity $ u_\infty $ indicates the calibrated wind tunnel velocity without the model and corrected by wind tunnel interference effects. The interference effects were corrected according to the methods proposed by \citet[][]{mercker1997contemplation} and \citet[][]{mercker1996correction}. The correction factors $ c_q = 1.01 $ for the dynamic pressure and $ c_u = 1.005 $ for the velocity were determined through this method. The correction factors are used in the sense of global values for the reference velocity $ u_\infty $ and for quantitative comparisons of the base pressure. A local correction would be necessary for field data. However, this is only necessary for quantitative comparison of the absolute values and is therefore not used here.

\subsection{Pressure measurements}
Time-resolved measurement data of the effects of the coherent structures on the base were recorded using pressure probes. The black dots in Figure \ref{ExpSetup} at the base mark the pressure measurement points. The base pressures were measured at $24$ points using the pressure transducers marked in Figure \ref{ExpSetup}. The pressure transducers were placed at coordinates $ y / h = \{ -0.58, -0.29, -0.04, 0.04, 0.29, 0.58 \} $, and $ z / h = \{ -0.33, -0.13, 0.17, 0.38 \} $. Each pressure transducer has an internal diameter of $ d_a = 0.7$ mm and is connected to an Esterline ESP-64HD pressure transducer via a $0.5 \hspace{5pt} m$ long pipeline with an internal diameter of $1.4 \hspace{5pt} mm$. The pressure transducers are calibrated and have a temperature compensation. The operating range of the pressure transducer is up to $6.9 \hspace{5pt} kPa$ with an accuracy of 0.05 \%. The sampling rate is $ f_s = 250 \hspace{5pt} Hz $ (sampling dimensionless frequency or Strouhal number $ S_s =  \{ 2.5, 1.875, 1.5, 1.25, 1.07 \} $). The resulting maximum frequency is expected to be at least capable of detecting vortex shedding signals (if present at this location) which are expected to be in dimensionless frequencies between $ \Delta Sr_h =0.1 $ and $ \Delta Sr_h = 0.25 $ (compare section 1). In order to take dynamic effects of the system of pipeline and measuring volume of the pressure transducer into account, a dynamic correction of the measured pressure values according to \citet[][]{bergh1965theoretical} was applied. Each pressure transducer was recorded over a total duration of 90 seconds. This corresponds to about 22103 measured values or time instants per pressure transducer and a dimensionless duration of $ t u_{\infty} / h = \{9000, 12000, 15000, 18000, 21000\} $ convective units. These timescales are expected to cover the switches of bi-stable behaviour (see section 1). Pressure readings are represented as dimensionless values $ c_p $ by the pressure difference to the plenum pressure $ p_\infty $ of the wind tunnel related to the dynamic pressure $ q_\infty = 0.5 u_\infty$. Time series spectra were used as Power Spectral Densities (PSD) with a Chebyshev window of size 150 and 50 \% overlap. This results in a frequency resolution of $ \Delta f = 1.66 \hspace{5pt} Hz$, or $ \Delta Sr_h =  \{ 0.0166, 0.0125, 0.01, 0.0083, 0.0071 \} $.  

In order to obtain time-resolved data of the flow field, additional point measurements were carried out in the shear layer at a Reynolds number of $ Re_h = 1.0 \hspace{5pt} 10^6 $. For this purpose, a multi-hole cobra probe of the manufacturer TFI with a scanning frequency of $ f_s = 3500 \hspace{5pt} Hz $ ($ Sr_s = 21$) was used to measure for 90 seconds ($ t u_{\infty} / h = 15 \hspace{5pt} 10^3 $ convective units). Thus the time-resolved velocity components were measured at 24 measuring points in the shear layer at the positions $ x / h = \{ 0.02, 0.17, 0.32, 0.49, 0.66, 0.83, 1.0, 1.17, 1.34, 1.51, 1.68, 1.85 \} $ with the corresponding coordinates $ z / h = \{ 0.51, 0.51, 0.51, 0.50, 0.49, 0.47, 0.43, 0.40, 0.36, 0.33, 0.33, 0.33 \} $ in the upper shear layer behind the separation edge. The PSD was used for these signals with a Hanning window of size $512$ and $50$ \% overlap. This leads to a frequency resolution of $ f = 3.4 \hspace{4pt} Hz $, or $ Sr_h = 0.02 $. The resulting maximum frequency for the shear layer measurements are expected to be at capable of resolving KH vortex signals (if present) after paring, e.g. in regions where the shear layer thickness is thick enough. For example for a shear layer thickness of $\delta_S = 0.3 h$ a dimensionless frequency of $ \Delta Sr_h =1.933 $  (compare section 1) or less is expected. This should be sufficient to follow at least to some extend the decrease of frequency through the vortex pairing along the growing shear layer.

\subsection{Particle Image Velocimetry}
Particle Image Velocimetry (PIV) was used to observe the flow around the near wake. For this purpose a Nd:YAG laser with a fixed frequency and thus a sampling rate of $10.5 \hspace{5pt} Hz$, a wavelength of $532 \hspace{5pt} nm$, and a maximum energy of $850 \hspace{5pt} mJ$ per pulse was used. The green areas or borders in Figure \ref{ExpSetup} represent the laser light sheet and the parallel staggered green areas the light sheets in the $y$-plane around the wake of the model. The thickness of the light sheet was about $2 \hspace{5pt} mm$. The incoming flow was mixed with aerosol droplets of Di-Ethyl-Hexyl-Sebacat (DEHS) with a diameter of about  $1 \mu m $. With two cameras of the type Imager sCMOS with a resolution of $2560 \times 2160$ images were recorded in a stereo PIV configuration. Thus an image area is available from two views from which all $3$ velocity components are determined. The cameras were arranged at an angle of $-20$\degree / $11$\degree to the $x$-axis in the $z$-plane, with a distance of $2 \hspace{5pt} m$ to the light sheet plane and equipped with a $50 \hspace{5pt} mm$ lens. The resulting field of view was $175 \times 140 \hspace{5pt} mm$. The time delay of the double images was $22 \mu s $. For the temporal averaging $300$ snapshots were recorded. The velocity vectors were calculated with an interrogation window of $32 \times 32$ with $75$ overlaps. The resolution is $1.8 \hspace{5pt} mm$ or $0.006 h$. No smoothing or filter was used in the evaluation. Several parallel $y$-planes of the wake were measured \citep[see][]{BockDiss}. In this publication only the $y = 0$ plane is considered. 

The sample rate for PIV is lower than the frequencies of interest for the vortex shedding and the KH vortices. For low frequency dynamics the measurement time is too short. Therefore PIV results are considered here for the structures of dominant fluctuating motions, based on a statistical (since not time resolved) representation of the wake dynamics. To observe the frequencies of the wake dynamics pressure measurements in the wake and at the base are considered. Complementary, simulation results provide a basis for combined resolved frequencies and dynamic structures for the modes of interest.

\section{Simulation setup}
The measurement data were extended by data from a simulation using the commercial software package SIMULIA PowerFLOW\textsuperscript{\textregistered} version 4.3d, which is based on the Lattice-Boltzmann method. Turbulence is treated using the Very Large Eddy Simulation (VLES) approach. The unresolved turbulent scales are represented by a variant of the RNG k model \citep[][]{yakhot1986renormalization}. The tangential velocity component on all friction walls is approximated with an extended standard model that includes the influence of the pressure gradient and surface roughness \citep[][]{launder1974numerical}.

\subsection{Flow field resolution}
The domain of flow simulation consists of a box around the SAE model. The domain covers a volume of $47 h \times 45 h \times 74 h$ ($14.1 \hspace{2pt} m \times 13.5 \hspace{2pt} m \times 22.1 \hspace{2pt} m$). The coordinate system is defined according to the experimental setup so that the $x$-axis runs parallel to the flow and the $z$-axis runs parallel to the height direction with the origin in the middle of the base surface. The inlet boundary condition with the defined flow velocity of $ u_{\infty}~=~50 \hspace{5pt} m s^{-1} $  (corresponds to $ Re_h = 1.0 \hspace{5pt} 10^6 $) is $20.6 h$ ($6.2 \hspace{5pt} m$) ahead of the leading edge of the model. 
As outflow boundary condition a constant pressure of $ p_{\infty} = 101300 Pa $ is imposed. A boundary condition of a friction-less wall (or also symmetry boundary condition) was applied to the lateral and the upper boundary of the domain. The vehicle itself is represented by a friction wall boundary condition in the simulation. The discretization is realized by a mesh of cubic simulation voxels. In $10$ resolution domains ($rd$), the resolution is halved from one level to the next in relation to the vehicle. The coarsest step has a resolution of $rd / h = 2.048$. A surface on the floor corresponding to the centerbelt was simulated as a friction wall surface with roughness moved along with the inflow $ u_{\infty} $. In offsets of the vehicle surface of \{ 0.352 h , 0.176 h, 0.088 h, 0.044 h, 0.022 h \} the resolution levels of $rd / h = \{ 0.032, 0.016, 0.008, 0.004, 0.002 \}$ were defined. The separation edge and the wake of the models were resolved with $rd / h = 0.004$ in a domain of $2.5 h \times 1.11 h \times 1.1666 h$ starting with $x = -0.066$ h from the base, with the ground and symmetrically around $y$. Around the A-pillars of the model, a further offset area by $0.044 \hspace{5pt} h$ was resolved with $rd / h = 0.004$. The simulation model comprises 47 million equivalent cells of the finest resolution level and is executed at an accelerated time step of $ \Delta t = 1.7 s $. The simulation was simulated for $ t u_{\infty} / h = 1043 $ convective units (corresponds to $6.26$ seconds) and was initialized with the solution of a previous simulation. Detection of the end of the simulation start-up process was performed according to the method proposed by \citet[][]{Mockett} based on the time series of the drag coefficient. The settling time is $ t u_{\infty} / h = 16.2 $ convective units (corresponds to $0.0969$ seconds). The remaining simulation time for averaging is $ t u_{\infty} / h > 1027 $ convective units (corresponds to $6.16$ seconds).
In \citet[][]{BockDiss} time-averaged data from measurement and simulation can be compared. The qualitative agreement of the distribution of velocities and velocity fluctuations in the wake, as well as the base pressures and the base pressure fluctuations is very good.

\subsection{Data acquisition}
Synchronized recordings of the base pressure and the $u$, $v$, $w$ velocity components (corresponding to $x$-, $y$-, $z$-direction) as well as the total pressure $ p_t $ of the flow field are used to observe the coherent structures. The wake volume was recorded with a spatial resolution of $rd / h = 0.00833$ (2.5 mm). In order to be able to look at two different time scales, two different data sets were recorded. The data set containing the high frequency time scales was sampled at a rate of $ f_s = 100 \hspace{4pt} Hz $ ($ Sr_h = 0.6 $) with a moving average over this period for a duration of $ t u_{\infty} / h = 333 $ convective units ($2$ seconds) and thus comprises $200$ snapshots of the volume. This recording is capable to resolve time scales of dimensionless frequencies between $ \Delta Sr_h =0.1 $ and $ \Delta Sr_h = 0.25 $  expected for the vortex shedding (compare section 1).

The low-frequency scales were recorded at a rate of $ f_s = 30 \hspace{4pt} Hz $ ($ Sr_h = 0.18 $) with a moving average over this period for a duration of $ t u_{\infty} / h = 1000 $ convective units (6 seconds). This data set thus contains 180 snapshots of the volume. Moving average is a filter of higher frequency time scales and thus of smaller length scales. This enables a clear representation of large-scale coherent structures. For the calculation of the PSD, a Chebyshev window of length $40$ with $50$ \% overlap, resulting in a frequency resolution of $ f = 2.5 \hspace{4pt} Hz $ ($ Sr_h = 0.015 $) was used for the high-frequency sampled data. For the low frequency sampled data a Chebyshev window of length $30$ with $50$ \% overlap resulting in a frequency resolution of $ f = 1 \hspace{5pt} Hz $ ($ Sr_h = 0.06 $) was used.

\section{Results}
In the presented results, the coherent structures in the postprocessing of an SAE vehicle model from the measurement and simulation data are described and documented using the methods mentioned above. The results refer to the Reynolds number $ Re_h = 1.0 \hspace{5pt} 10^6 $, whereby the same representations of the experimental data also apply to all other Reynolds numbers investigated. The results will be presented in the following order. At the beginning, the averaged flow field is observed. Some special features of the wake topology, i.e. the vortex ring, are highlighted in comparison to other documented vehicle models. The subsequent considerations of the base pressure distribution and the motion of the barycenter of pressure give first insights into the areas with the strongest energy losses (in the form of low base pressures) and the directions of motion of the wake. The frequencies of the dimensionless, spatially averaged surface pressure gradient (equation \ref{Grad}) and the velocity fluctuations in the shear layer give an impression of the time scales of the motions in the wake. Thereupon the motions in the middle frequency range of the wake in the flow field are considered. In this frequency range a vortex separation process is to be expected, which is then described in more detail with this observation. Finally, the low frequency motions of the flow field in the wake are considered.

\subsection{Description of the time averaged near wake flow field}
\begin{figure}[h]
  \centerline{
  \includegraphics[width=12cm]{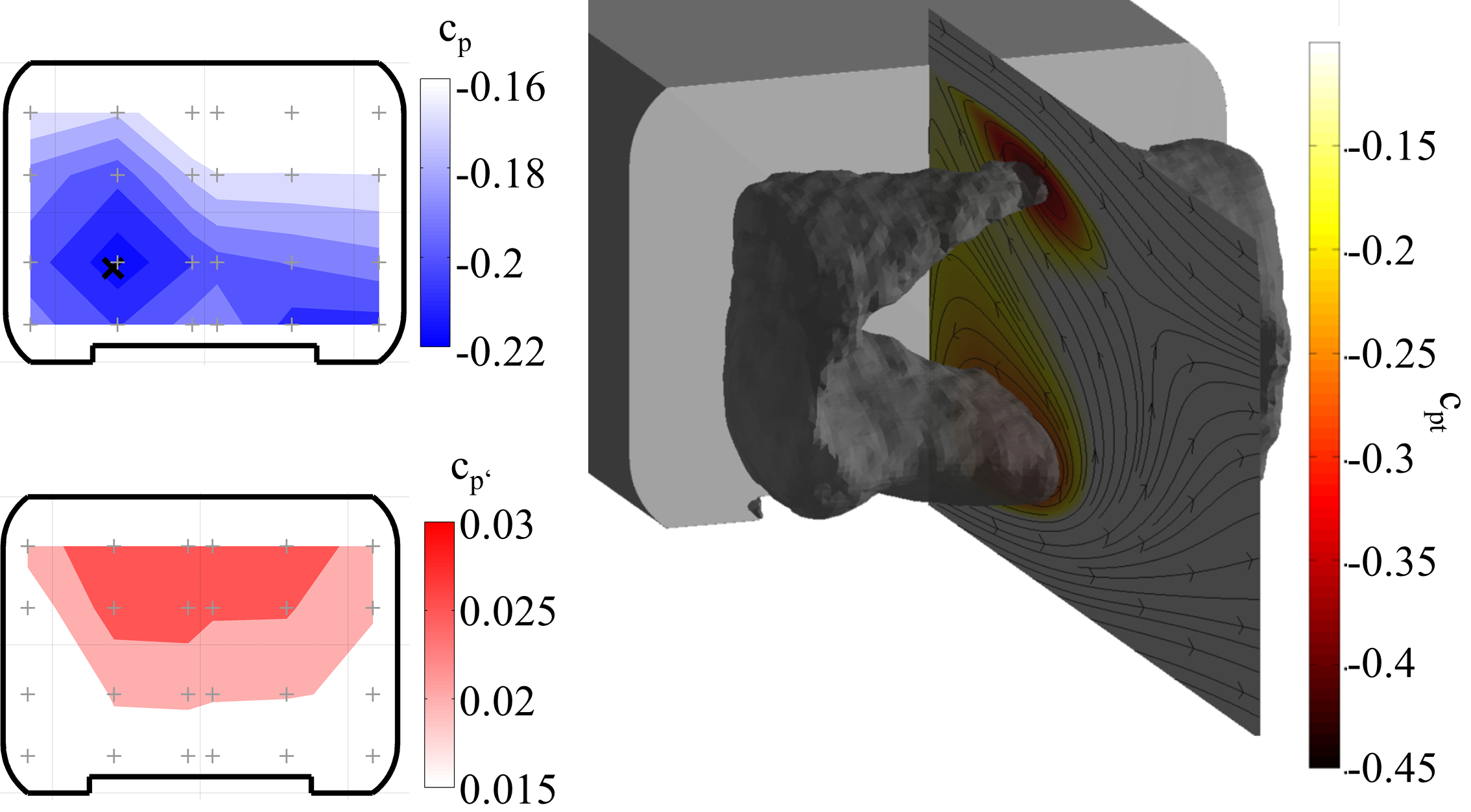}
  }% Images in 100% size
  \caption{Averaged base pressure (left top), base pressure fluctuations (left below) from experiments and visualization of the wake flow around the SAE squareback model with isosurface of $c_{p,t} = -0.23$ from simulation}
\label{WakeAv}
\end{figure}

The dead water of the SAE fullback model consists of a vortex ring which is deformed  in the upper section of the dead water by the influence of the diffuser and moves closer to the base in the lower section. The shape of the vortex ring and thus the dead water area differs from the shape of the vortex ring on an Ahmed squareback model. Figure 2 describes on the right hand side the flow volume in the wake based on the simulation data. The right figure shows the rear end of the model (in grey) and the streamlines in the middle plane of the wake with coloring of the total pressure coefficient, as well as the isosurface of the total pressure $ c_{pt} = -0.23 $. The streamlines outline the recirculation area with the vortex ring. The simulation results allow the additional consideration of total pressure as an indicator for irreversibly converted kinetic energy, i.e. contributions to drag. The total pressure values are particularly low in areas around the vortex ring. As expected, high contributions to the aerodynamic drag are localized around the vortex ring. The selected isosurface can be used to describe the shape of this vortex in the volume of dead water. The upper part of the vortex ring grows in the y-direction from the center outwards. In addition, a slight deformation can be observed in the lower part of the vortex ring, in the section where the diffuser ends. These characteristics of the dead water area represent the most conspicuous differences to the wake of other squareback models \citep[compare][]{BockDiss, Grandemange2013, duell1999experimental, Barros2015, Grandemange2013a, brackston2016stochastic, khalighi2001experimental}, none of which has a diffuser.
The left part of Figure \ref{WakeAv} shows the distribution of the base pressures at the top and their fluctuations from the measurement data at the bottom in a blue-white or a red-white coloring respectively. The same distributions result qualitatively from the simulation data (not shown here, \citep[compare][]{BockDiss}). The illustrations outline the base surface in plan view with a thick line, as well as the positions of the pressure measuring points with grey crosses. In the display of the base pressures, the barycenter of pressure is also marked as black x. The lowest base pressures correlate to the local proximity and size of the vortex ring in the lower part. The pressure distribution is skewed. Both of these properties are represented by the position of the barycenter of pressure. The largest fluctuations of the base pressures, however, are located in the upper half of the base and around the middle in the y-direction.

\subsection{Spatial distributions of variances of base pressure distribution}
\begin{figure}[h]
  \centerline{
  \includegraphics[width=12cm]{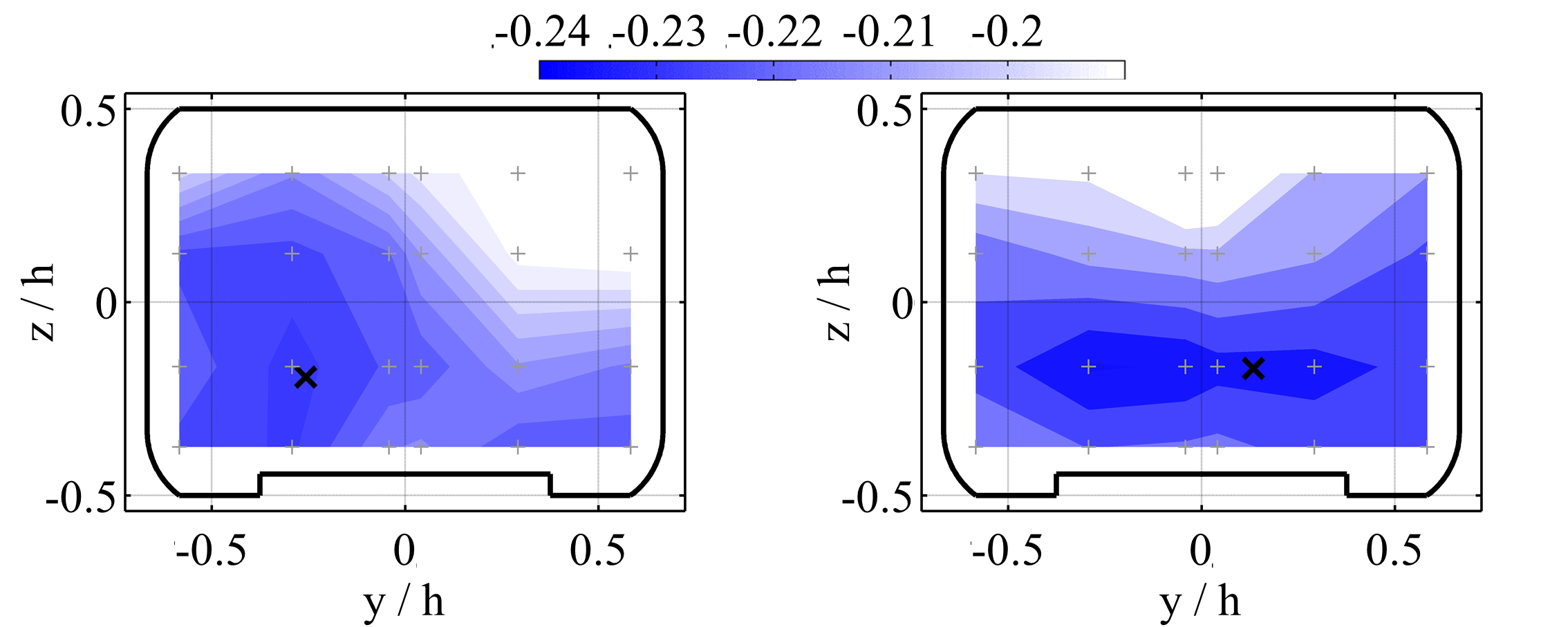}
  }% Images in 100% size
  \caption{Snapshots of base pressure distributions from experiments}
\label{SPp}
\end{figure}

Snapshots of the pressure distribution show mainly deflections of the distribution in the width and the lowest pressures in the lower range in $z$-direction. The motions of the wake can be analyzed with the distribution of the base pressure. The higher sampling rate of the pressure measurements compared to the flow field measurements can be used as an advantage in terms of temporal resolution of coherent structures. Figure \ref{SPp} illustrates some basic properties of this observation in the form of two snapshots. The images of the two snapshots show the $y$-direction (width) above the abscissa and the $z$-direction (height) above the ordinate. The separation edge is shown as a black thick line. Grey crosses mark the pressure measuring points. The pressure distribution is shown by the coloring. A barycenter of the lowest pressure is calculated for the respective distribution of the time instant and displayed as a black $ \times $ to demonstrate how the determined pressure center corresponds to the pressure distribution. The left snapshot shows an asymmetric distribution. In principle, the pressures in the lower ranges ($ z / h < -0.4 $) are low for both snapshots. At the asymmetric distribution (left side) pressures at right upper quarter are high. The symmetrical distribution shows the highest pressures in the upper, middle area, which decreases slightly outwards and slightly more downwards. The calculated focal points correspond well with the visual distribution and are therefore suitable for further evaluations.

\begin{figure}[h]
  \centerline{
  \includegraphics[width=8cm]{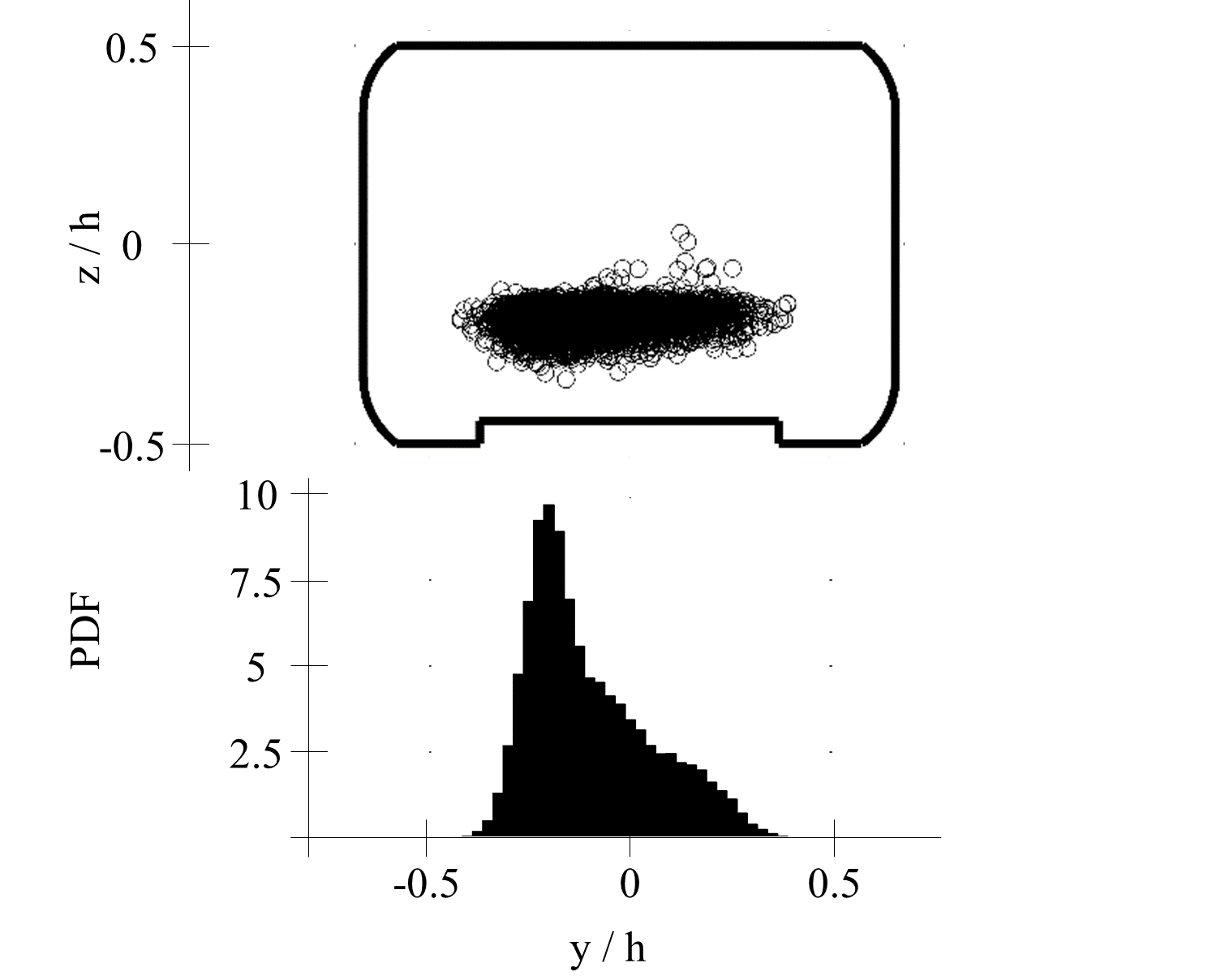}
  }% Images in 100% size
  \caption{Distribution of barycenters of pressures from snapshots on the base surface (top) and distribution function along y from experiments.}
\label{ScatterSP}
\end{figure}

The barycenter of the low pressures lies in the lower half and moves mainly in the width. Figure \ref{ScatterSP} shows the distribution of the barycenter positions from snapshots of the pressure distribution. The rear edge is marked with a thick black line, the $y$-direction is plotted on the abscissa and the $z$-direction on the ordinate. It is shown that the barycenter and thus also the wake moves primarily in the $y$-direction and comparatively little in the $z$-direction (i.e. height). It should be noted that the barycenter is in the lower half. However, the motions of the barycenter are due to the fluctuations in the upper half, as the fluctuations in this area dominate (see Figure \ref{WakeAv}).

The distribution of the snapshots of the barycenters of the base pressure distribution indicate a lateral deflection of the wake and thereby exclude a bi-stability. The distribution of the pressure points of the individual snapshots confirms these findings. The pressure centers are represented by rings with the indicated rear contour in Figure \ref{ScatterSP} above. Most of the barycenters from the series of 5000 snapshots are located in the lower half of the base. The points of these snapshots correspond to $ t u_{\infty} / h = 3300 $ convective units which should be long enough to experience a bi-stable switch of lateral deflection. Even considering the complete series of 22103 snapshots ($ t u_{\infty} / h = 15000 $) the distribution of pressure centers do not change. Instead, the focal points are distributed over a large span along the width. From the shift of the barycenter it can be concluded that the wake is deflected more in $y$- than in $z$-direction. Figure \ref{ScatterSP} below shows the frequency distribution of these positions along the width. The distribution of the barycenters shows an obliquity. However, the focus is clearly on one position and not two. A bi-stable behavior can thus be excluded as the cause of the lateral deflections of the wake.
\subsection{Time scales of variances of base pressure distribution}
\begin{figure}[h]
  \centering
  \includegraphics[width=12cm]{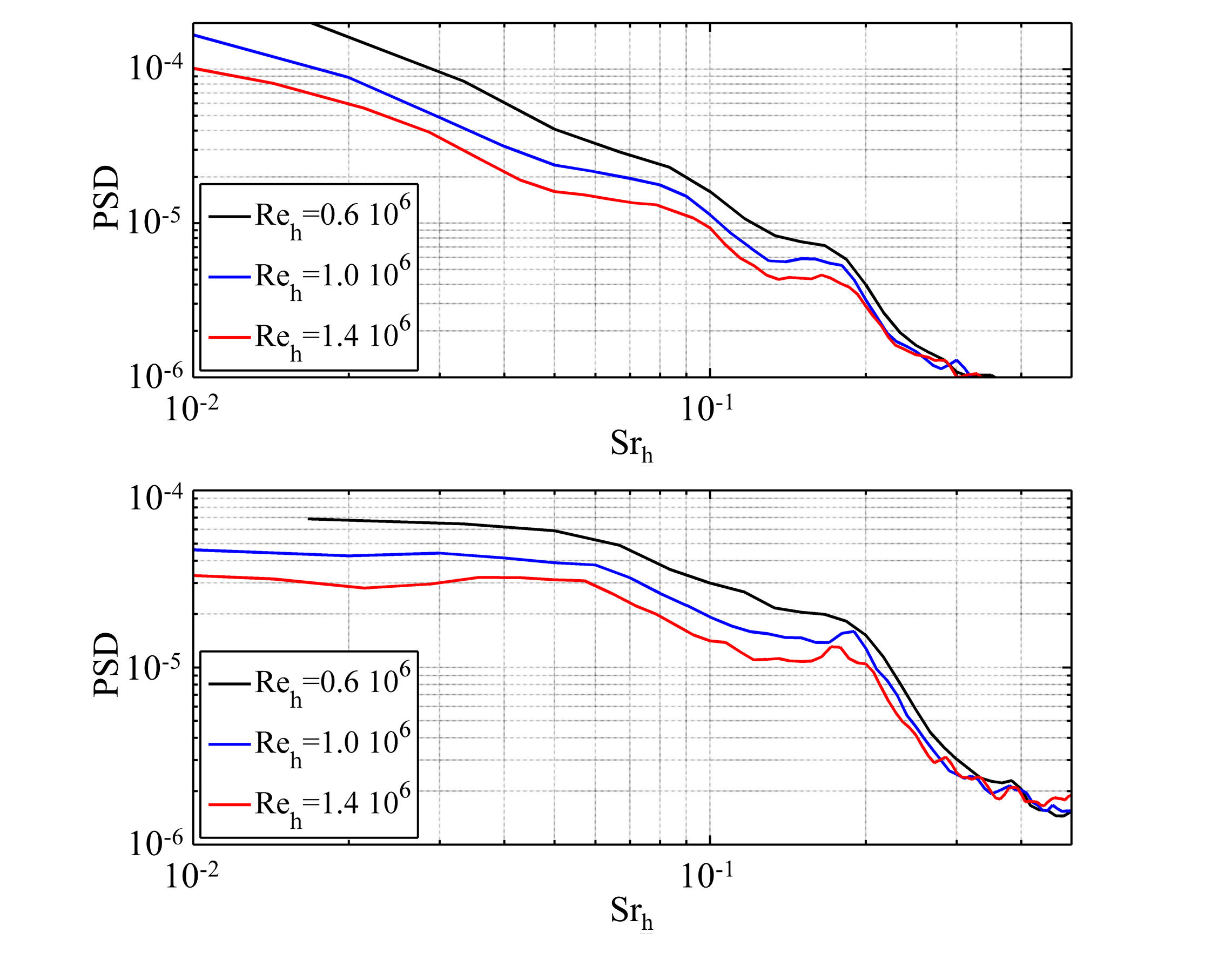}
  \caption{Spectra of the dimensionless, spatially averaged surface pressure gradient on the base surface in y- (top) and z- direction (bottom) at different Reynolds numbers from experiments.}
\label{SpektrenBasis}
\end{figure}

The base pressure fluctuations are mainly dominated by the effects of the deflection of the wake in the direction of the width with a low frequency. However, the effects of deflections in width and height direction in the medium frequency range can also be determined in the base pressures. Figure \ref{SpektrenBasis} shows the PSD of the time series of the dimensionless, spatially averaged surface pressure gradient in $y$- (top) and $z$- (bottom) direction for different 
Reynolds numbers. In the spectra, very low frequencies ($ Sr_h < 0.03 $) of the motions in y- and z- direction, but especially in y- direction, are dominant. For the y- position, a survey or at least a plateau at the frequency $ Sr_h \approx 0.16...0.18 $ is also recognizable for all Reynolds numbers. In z-direction, the barycenter of pressure is deflected less severe. Nevertheless, the fluctuations at $ Sr_h \approx 0.17...0.19 $ are also discernible in $z$-direction. All raised areas appear as very broadband distributed fluctuations. The strongest deflection occurs at low frequencies. However, there is also a deflection of the wake, which moves in both directions with time scales that correspond approximately to dimensionless frequencies of $ Sr_h \approx 0.16...0.19 $. The spectra of some individual pressure measurement points (not shown here) also show these frequencies \citep[see][]{BockDiss}.

\subsection{Time scales in the shear layer}
\begin{figure}[h]
  \centering
  \includegraphics[width=12cm]{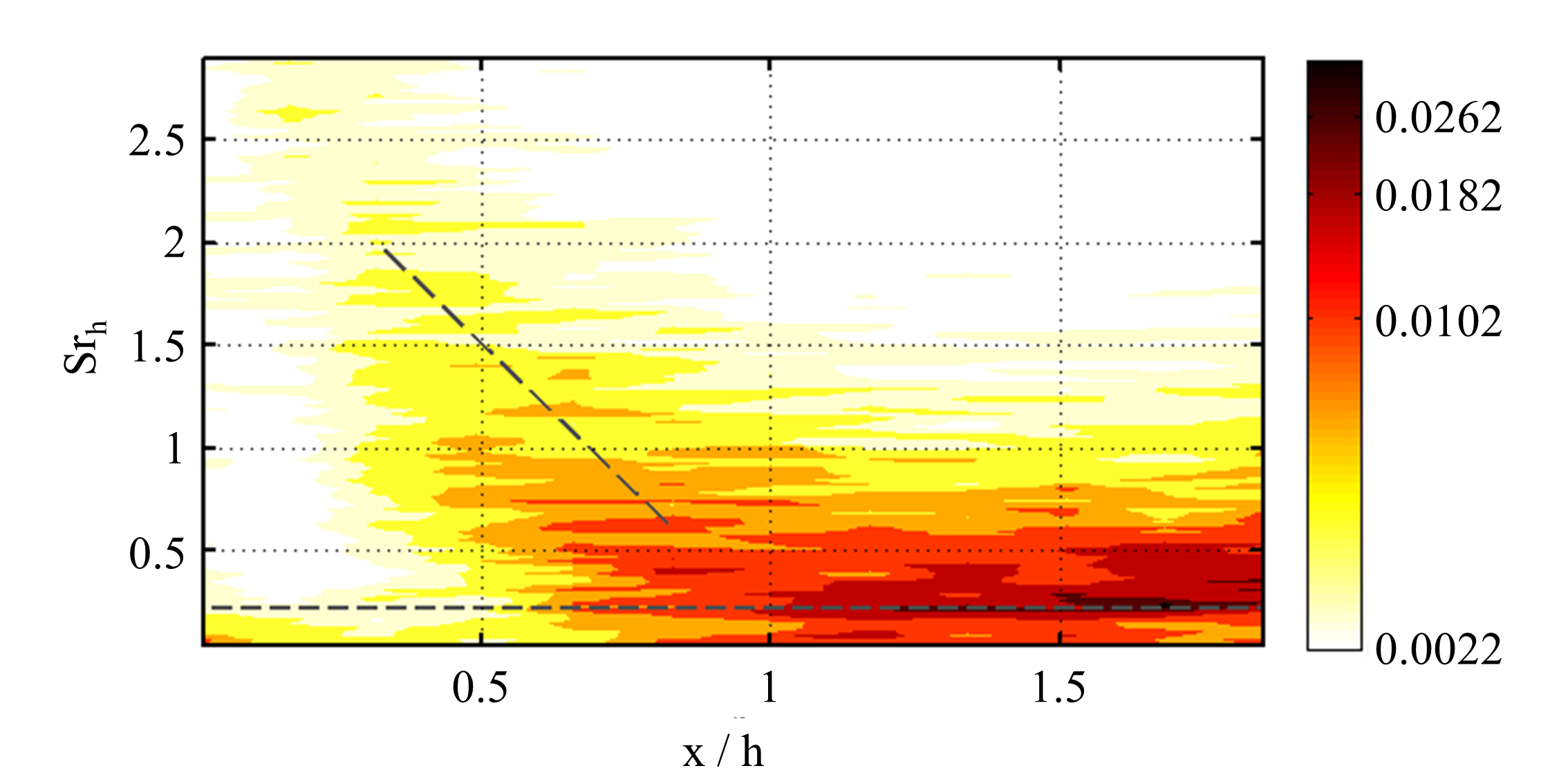}
  \caption{Spectra (fluctuating part of PSD: $ u_y'^2 $) over x-position along the upper free streamline in the wake plane $y/h=0$ from experiments.}
\label{fig:SpektrenScherschicht}
\end{figure}

In the vicinity of the separation edge, a decreasing frequency peak at high frequencies correlates with the increase in the shear layer thickness. Spectra from the time series of measuring points from the upper free streamline of the shear layer in the center plane ($y / h = 0$) are shown in Figure \ref{fig:SpektrenScherschicht}. The PSD of the $ u_y' $ fluctuating velocity component is plotted over different positions on the abscissa, while the dimensionless frequency $ Sr_h $ is plotted on the ordinate. The PSD is indicated by the coloring with logarithmic scaling. The fluctuations in the upper shear layer are highest (cf. Figure \ref{WakeAv}). For this reason, the lower shear layer is not shown here. However, they can be seen in the work of \citet[][]{BockDiss}. In principle, distributions of the fluctuations of the upper and lower shear layer are similar. A feature of these spectra is the shift of a broadband peak near the base from high frequencies to low frequencies further downstream. The first occurence is recognizable at $x / h = 0.35$ with $ Sr_h = 2 $, which changes up to $ Sr_h \approx 0.6 $ at $x / h = 0.8$. This transition is marked in the illustration with a dashed, diagonal line. This process describes a drop in the characteristic frequency $ f_c \sim u_0 / \delta_S $ in the direction of flow, which is also characteristic of free shear layers. In this relation $ \delta_S $ is the shear layer thickness. The velocity $ u_0 $ represents the time averaged flow velocity in the center line of the shear layer. This velocity undergoes only minor changes in the considered range of $x / h$ \citep[see][]{BockJWEI}. The almost linear decrease of the frequency $ f_c $ is caused by the rolling up and thus the growth of KH vortices in downstream flow direction in the shear layer. In the work of \citet[][]{BockJWEI} a linear increase of the shear layer with $ \delta_S / dx \approx 0.1 $ is given here. According to \citet[][]{champagne1976two} and \citet[][]{dimotakis1991turbulent} this also resembles the free shear layer with $ \delta_S / dx = 0.06...0.11 $. Similar values are given by \citet[][]{Grandemange2014c} with $ \delta_S / dx = 0.14 $, or $ \delta_m / dx = 0.12 $ for the vehicle types Renault Trafic, respectively Peugeot 3008. The behavior of this high-frequency peak is also consistent with the observations of \citet[][]{duell1999experimental} on a squareback model.

The strongest fluctuations in the shear layer mainly occur with $ Sr_h \approx 0.2 $ near the saddle point ($ x / h \approx 1.5...1.7 $). However, high levels of fluctuations in very low frequencies are strongly present over the entire shear layer. It is particularly noticeable in Figure \ref{fig:SpektrenScherschicht} that the fluctuations are highest in the range $ x / h > 0.8 $. This correlates with the observations of the fluctuations in the flow field in Figure \ref{WakeAv}. The broadband distributed fluctuations increase, with maximum distance to the base for example at $ x / h = 1.6 $ around a frequency of $ Sr_h \approx 0.22 $, which is drawn as a horizontal dashed line. The accumulation of fluctuations around $ Sr_h \approx 0.22 $ can still be seen further downstream.

\subsection{Intermediate frequency wake modes}
\begin{figure}[h]
  \centerline{
  \includegraphics[width=12cm]{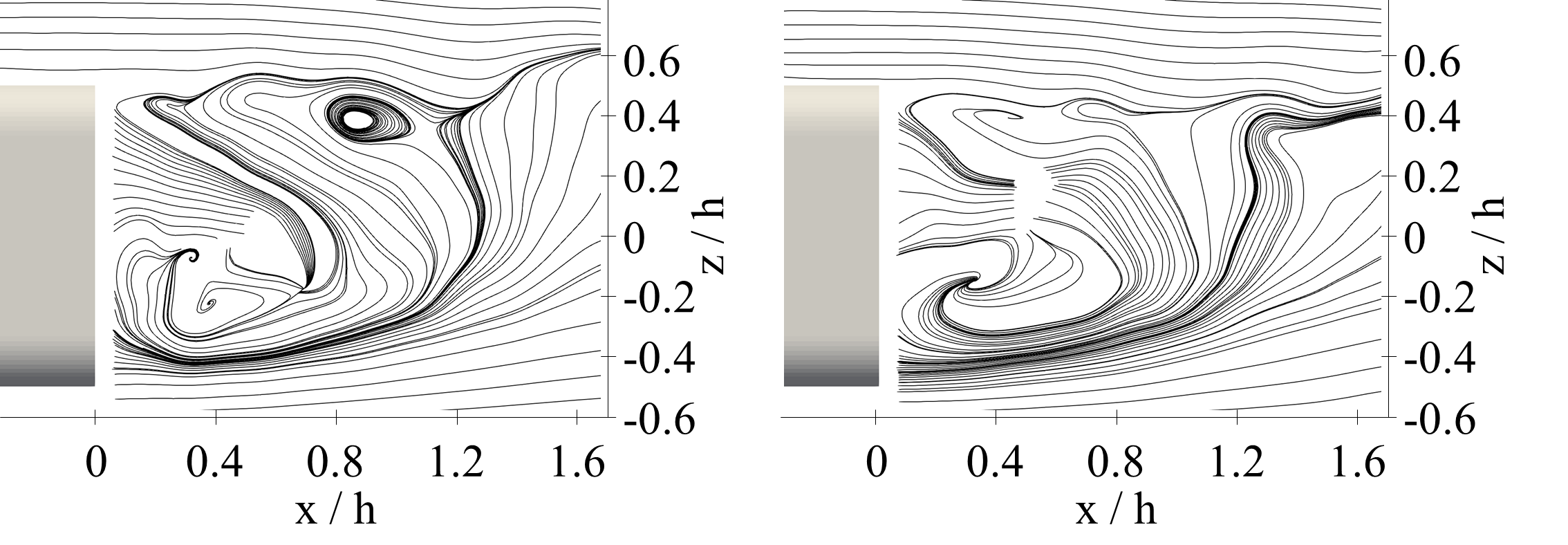}
  }% Images in 100% size
  \caption{Snapshots in the wake plane $y / h = 0$ from experiments.}
\label{fig:SnapsWake}
\end{figure}

In this work intermediate frequencies are considered as frequencies lower than frequencies of KH vortices, related to vortex shedding with vortices larger than the KH vortices. Snapshots of the dead water show vortex structures of different sizes, especially in the area of the upper shear layer close to the base. A first impression of the coherent structures results from a few snapshots of the flow field. As an example, Figure \ref{fig:SnapsWake} shows two snapshots experimentally determined for the center plane ($y / h = 0$). The illustration shows the rear edge of the model in grey and the streamlines in the plane of the wake. In the lower domain ($z / h < 0$), both snapshots show curved streamlines at different positions. In the upper domain ($z / h > 0$) curved streamlines are also recognizable. These are mainly located at the boundary with rather straight streamlines, i.e. in the shear layer. In the left part of Figure \ref{fig:SnapsWake}, there are curved (but not closed) streamlines in the upper section at $ x / h \approx 0.25 $ and $ x / h \approx 0.5 $. At $ x / h \approx 0.9 $ a vortex with a comparatively larger diameter can still be seen. These undulating, curved streamlines in the upper section are an indication of vortices that grow with the direction of flow. This correlates with the growth of the shear layers in the direction of flow and the associated characteristic vortex structures. On the right side of Figure \ref{fig:SnapsWake} two closed curved streamlines are shown in the upper area, i.e. vortices at $ x / h \approx 0.6 $ and $ x / h \approx 1.3 $. Both vortices rotate in clockwise direction and correspond to the expected direction of rotation in the upper shear layer. The streamlines starting from the lower and upper separation edge meet at the instantaneous position of the saddle point at about $ x / h \approx 1.0 $ in the upper domain. So the further downstream vortex is behind the saddle point and thus outside the dead water. This snapshot shows already detached vortex structures of the upper shear layer which are further transported into the far wake. 

\begin{figure}[h]
  \centerline{
  \includegraphics[width=12cm]{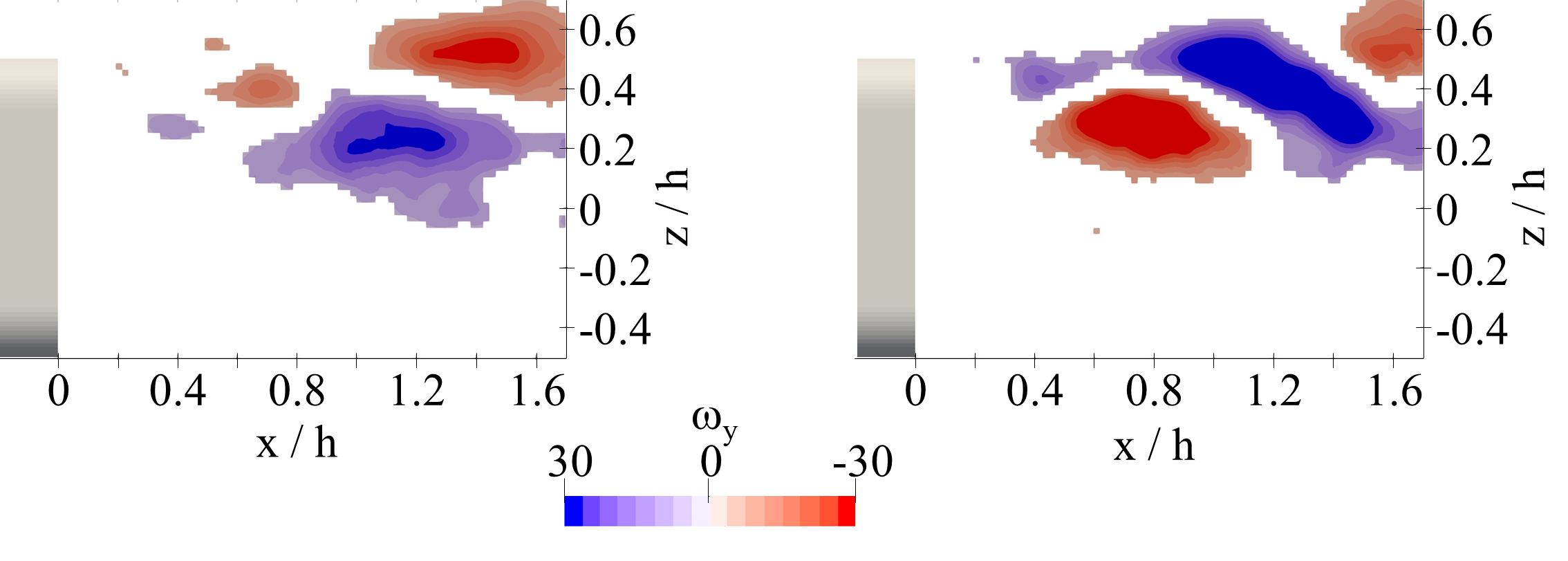}
  }% Images in 100% size
  \caption{Vorticity distribution of modes $m=1$ (left) and $m=2$ (right) in the wake plane $y / h = 0$ from experiments.}
\label{fig:ModenWake}
\end{figure}

A more detailed description of the processes and the spatial scales is enabled by the reconstruction from the phase references of the first mode pair of the POD. The first two POD modes of the flow field represent a convective mode pair and indicate a vortex separation process. Figure \ref{fig:ModenWake} shows the distribution of the vorticity in the wake of the SAE model in the $y / h = 0$ plane of the POD modes $m=1$ (left) and $m=2$ (right). The two POD modes of the flow field represent structures of the same size and wavelength, which are only spatially shifted by a quarter of a wavelength. These structures are also shown at other measured planes (not shown here, \citep[see][]{BockDiss}). Thus a convective mode pair can be assumed. A phase-averaged reconstruction based on the phase definition between this mode pair can be conducted (see equation \ref{phase}).

\begin{figure}[h]
  \centerline{
  \includegraphics[width=12cm]{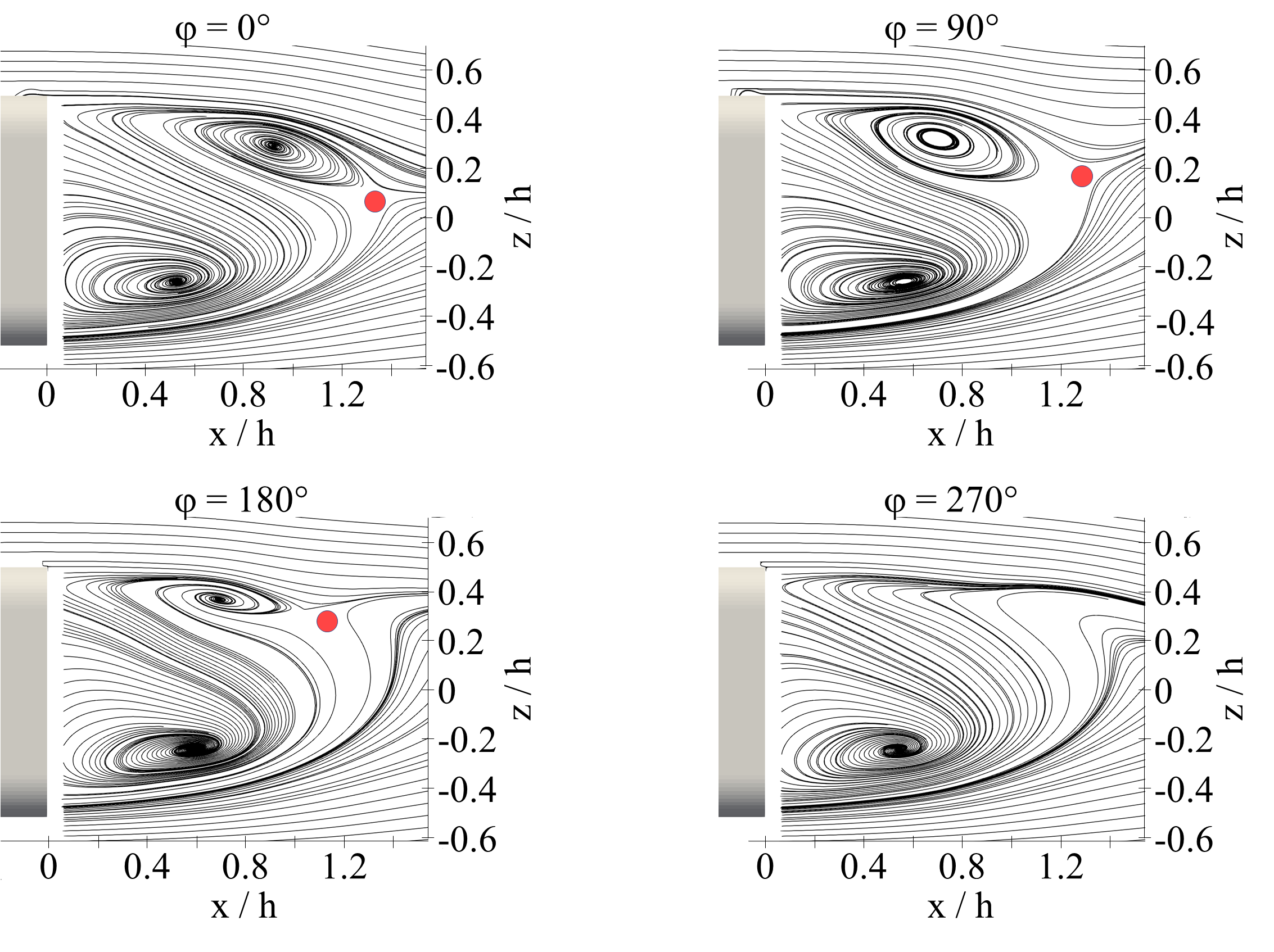}
  }% Images in 100% size
  \caption{Reconstruction from two modes superimposed with the averaged flow field in the $y / h = 0$ plane (phase offset of $ \Delta \varphi = 113^\circ$ based on the phase definition between amplitudes of the modes $m=1$, $m=2$) from experiments.}
\label{fig:RekoEbene}
\end{figure}

In reconstructions of POD of individual planes, a vortex motion occurs as a global mode in the upper dead water. The streamlines from the reconstructed flow fields can be seen in Figure \ref{fig:RekoEbene}. This reconstruction of the wake plane represents the flow in the center ($y / h = 0$) based on the phase definition with modes $m=1$ and $m=2$ from the POD. The phase position was shifted by $ \Delta \varphi = 113 \degree $ without affecting the generality of the results. The illustration shows streamlines and the rear of the model in grey. Starting with the phase $ \varphi = 0 \degree $ (above, left) vortices are visible in the dead water, which in principle resemble the arrangement of the averaged flow (Figure \ref{WakeAv} right side), with a larger lower vortex close to the base and a smaller upper vortex which is more downstream to the base. It is noticeable in this phase that the upper vortex is larger compared to the averaged flow. Connected to this observation, the position of the saddle point is slightly shifted downwards. The saddle point (in the figure marked by a red dot) is interpreted as an indicator for the length of the dead water region. In the following phase $ \varphi = 90 \degree $ the upper vortex becomes larger in height (z - direction) and its center moves towards the base. This shifts the saddle point upwards. At the same time, the streamlines behind the saddle point are deflected upwards. As in the following phases, the lower vortex hardly changes, at $ \varphi = 180 \degree $ the upper vortex becomes very small in the dead water. However, curved, but not closed streamlines can be seen behind the saddle point at $ x / h \approx 1.2 $. This indicates vortex structures downstream of the dead water. In phase $ \varphi = 270 \degree $ the upper vortex in the dead water can no longer be identified. The saddle point as convergence point of the streamlines of the upper and lower wake area must be either significantly closer to or significantly further away from the base surface than in the other phases. It should be noted that the lower free streamline (which separates the outer flow from the dead water) starts above $ z / h > -0.5 $, as the rear diffuser is located in this plane. At $ x / h \approx 1.2 $ the streamlines form non-closed curved paths. All this indicates a vortex shedding from the dead water. Meanwhile, the length of the dead water changes considerably and a motion of the saddle point or a fluctuation of the expansion of the dead water area develops in the direction of flow. 

In the work of \citet[][]{BockDiss} these processes are shown for further wake planes by mode structures and by streamlines of their reconstruction from experiments. It can be seen that the modes $m=1$, $m=2$ in the other planes are convective modes and that they behave very similarly, taking into account the fundamental change in the temporally averaged flow over the width. Nevertheless, with the evaluation of individual planes, ordered processes across the planes can hardly be considered. To illustrate the three-dimensional flow of the process in Figure \ref{fig:RekoEbene}, POD mode decomposition from the simulation is used below.

\begin{figure}[h]
  \centerline{
  \includegraphics[height=4.25cm]{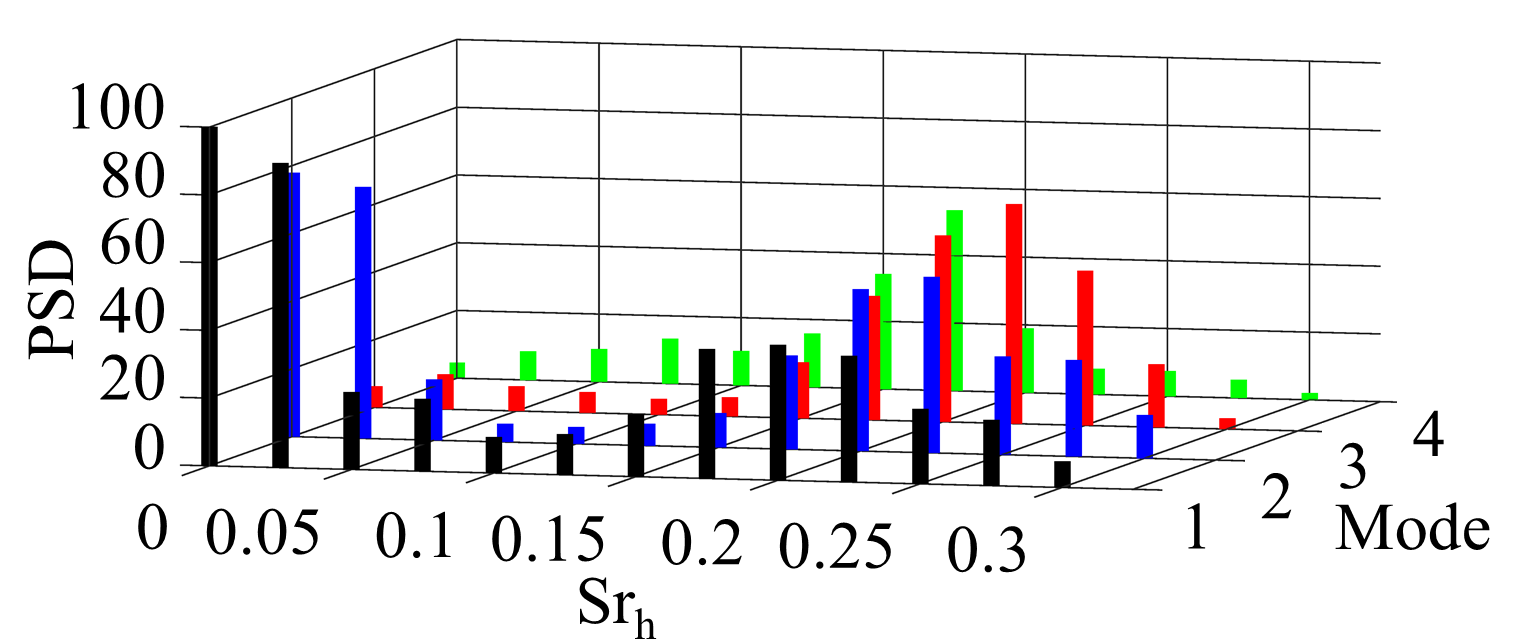}
  }% Images in 100% size
  \caption{Spectra (Fluctuating part of PSD: $ a_m^2 $) of amplitudes from POD in the wake volume from simulation.}
\label{fig:SpektrenPOD}
\end{figure}

\begin{figure}[t]
	\centering
           \begin{subfigure}[]{\textwidth}
              \centering
               \rule[1ex]{\textwidth}{0.4pt}
               \caption*{$ \varphi = 0 \degree $}
	    \includegraphics[width=12cm]{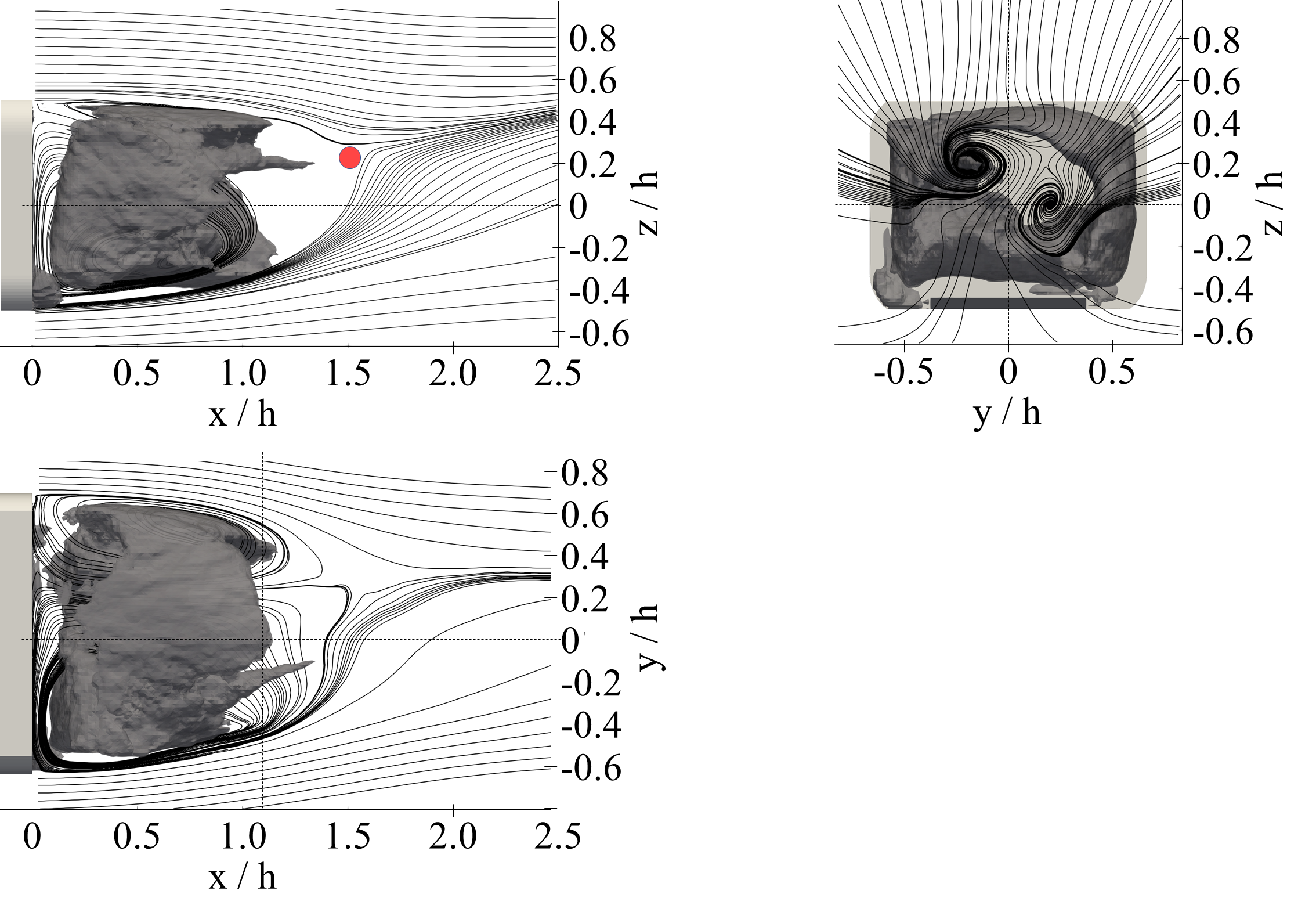}	 	
               \rule[1ex]{\textwidth}{0.4pt}
            \end{subfigure}
            \begin{subfigure}{\textwidth}
              \centering
               \caption*{$ \varphi = 90 \degree $}
	    \includegraphics[width=12cm]{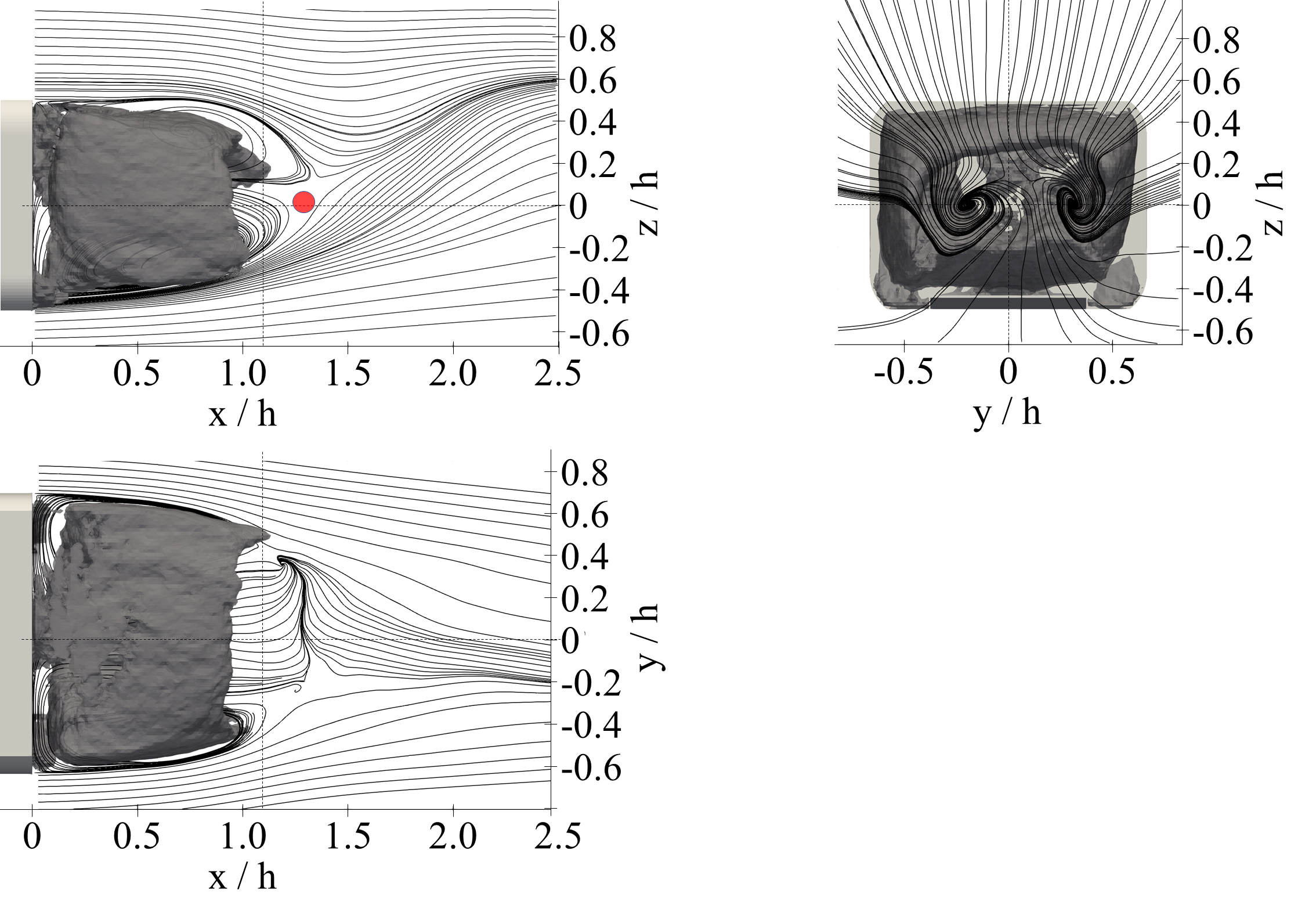}   
               \rule[1ex]{\textwidth}{0.4pt}
	  \end{subfigure}
             \caption{Reconstruction from the first $4$ POD modes in superposition with time averaged flow. Isosurfaces of $ c_{p,t} = -0.1 $ from simulation.}
\label{fig:RekoVol1}
\end{figure}

\begin{figure}[t]
	\centering	
           \begin{subfigure}[]{\textwidth}
               \centering
               \rule[1ex]{\textwidth}{0.4pt}
               \caption*{$ \varphi = 180 \degree $}
               \includegraphics[width=12cm]{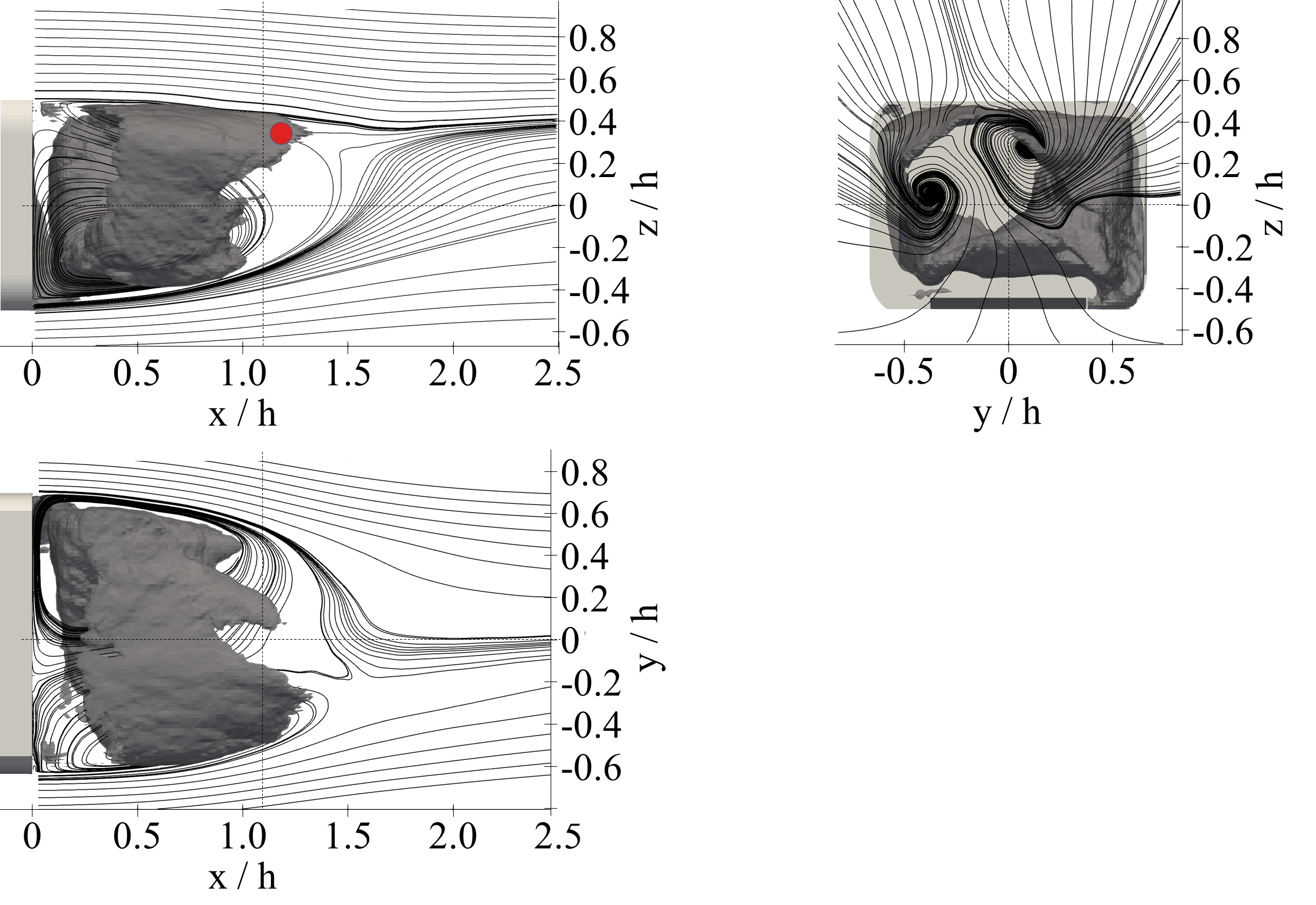}	 	
               \rule[1ex]{\textwidth}{0.4pt}
	  \end{subfigure}
           \begin{subfigure}[]{\textwidth}
               \centering
               \caption*{$ \varphi = 270 \degree $}
               \includegraphics[width=12cm]{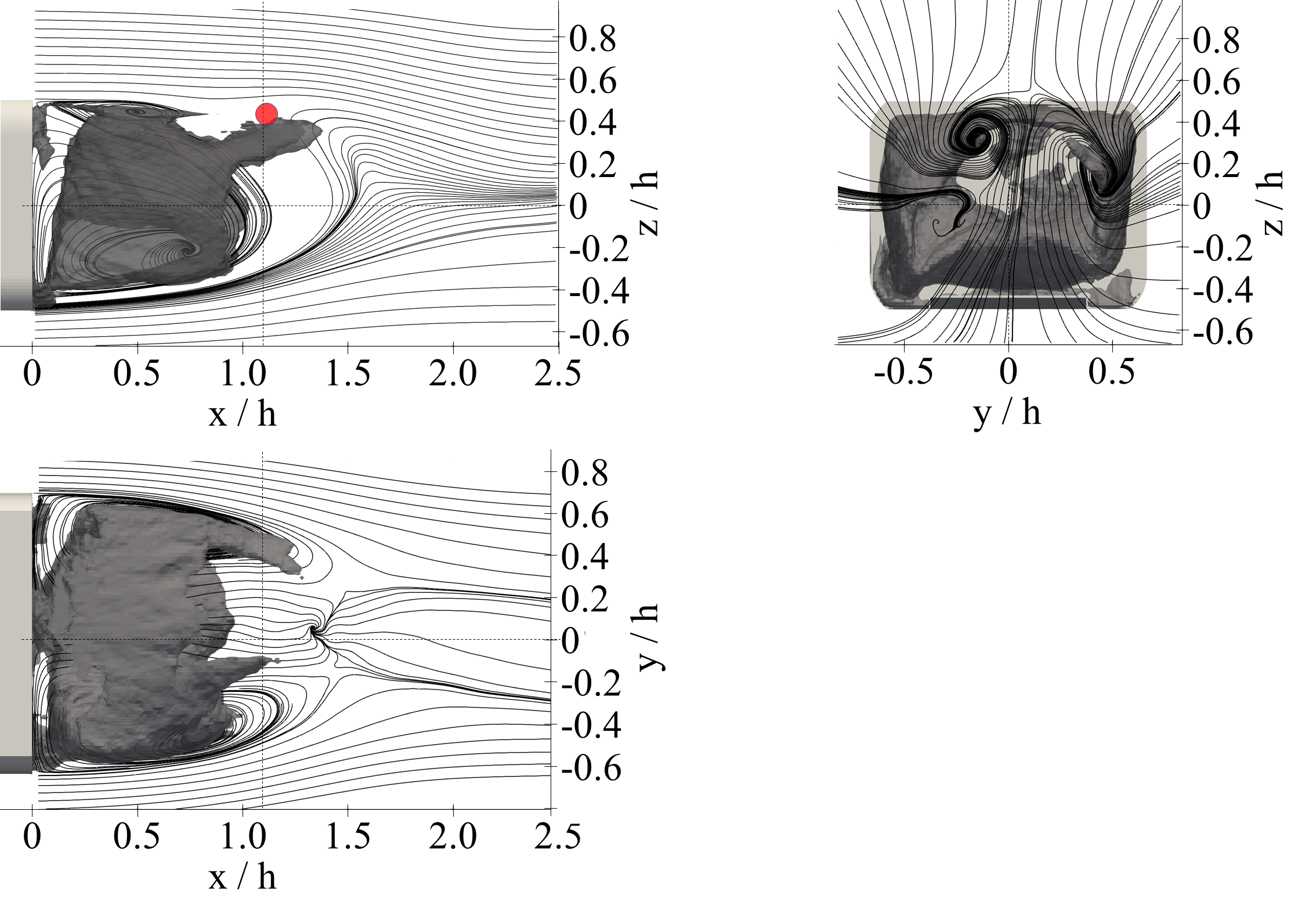}   
               \rule[1ex]{\textwidth}{0.4pt}
	  \end{subfigure}
	  \caption{Reconstruction from the first $4$ POD modes in superposition with time averaged flow. Isosurfaces of $ c_{p,t} = -0.1 $ from simulation.}  
\label{fig:RekoVol2}
\end{figure}

The POD modes from $m=1$ to $m=4$ of the high frequency flow volume recording from the simulation represent two convective mode pairs that are temporally and spatially paired. The behavior of the dominant fluctuation motions in the context of the considered mean time scales can thus be modelled from these $4$ modes. The composition of POD modes in the wake volume differs in some small details from the POD modes of the plane flow field from the measurement. This can be due to the different sampling rates, but also to the observation of the correlations in the flow field over the entire volume instead of just one plane. However, the structure of POD modes $m=1$ and $m=2$ of the flow volume from simulation is similar to Figure \ref{fig:ModenWake}. Hence they represent a convective mode pair. In the same way, mode $m=3$ and $m=4$ describe a convective mode pair. Furthermore, the modes of the flow field in the $y$-plane demonstrate that the simulation results show the same mode structure with the same magnitude scale as the experimentally determined modes. The amplitudes spectra of POD modes from $m=1$ to $m=4$ shown in Figure \ref{fig:SpektrenPOD} show very dominant peaks in the same frequency range. Mode from $m=1$ to $m=4$ have the most dominant fluctuating parts in the frequency band around the broadband peak of $ Sr_h \approx 0.22 $. Modes $m=1$, $m=2$ also have strong fluctuations at low frequencies. This constellation of convective mode pairs with the same dominant frequencies justifies a reconstruction with the first $4$ POD modes by a modeling based on the phase definition between the first mode pair (see equation \ref{phase}). In terms of the motions of the upper wake vortex part, the wake saddle point  and the holding position of the lower vortex  in the $y / h = 0$ plane (Figure \ref{fig:RekoVol1} to Figure \ref{fig:RekoVol2}), this modeling shows the same behavior as in the experiment (Figure \ref{fig:RekoEbene}) and therefore reinforces this approach.


In the near wake volume, the global mode is represented by the vortex shedding with a superimposed, global deflection of the wake in direction of the vehicle width. Figures \ref{fig:RekoVol1} and \ref{fig:RekoVol2} show the reconstruction of the process with $ Sr_h \approx 0.22 $ in the wake volume using four phases. In the pictures, divided into $2$ rows, $2$ phases with $3$ different views are shown. These views of the $y$- (top, left), $x$- (top, right) and $z$- (bottom left) planes are intended to illustrate the three-dimensional motion of the wake volume. In the views, the rear of the model is shown as a light gray geometry. An isosurface of total pressure $ c_{p,t} = -0.1 $ is visualized to trace the motions of the dead water. In order to facilitate the characterization of the dead water area the location of the saddle point in the $y$-plane is marked with a red point. The saddle point is interpreted as an indicator of the length of the dead water region. To support the interpretation of the motion, the streamlines are shown in a plane in each view. The positions of these planes ($y / h = 0$, $x / h = 1.1$, $z / h = 0.35$) are marked as dashed lines in the orthogonal views. The phase states can be described as follows:

$ \varphi = 0 \degree $: In the first row of Figure \ref{fig:RekoVol1}, the view of the y-plane is shown at the top and left. In the upper section of the dead water a vortex is located at $ x/h \approx 0.5 $. The saddle point in the plane of the streamlines is located in the upper half of the wake at $ x / h \approx 1.5 $, $ z / h \approx 0.2 $. In the x-plane an asymmetric situation emerges. The streamlines of the x-plane show two vortices with their focuses located at $ z / h \approx 0.2/0 $, $ y / h \approx +/-0.2 $. Accordingly, the total pressure in this plane is distributed asymmetrically. On inspection of the streamlines in the dead water in the z-plane this asymmetry is also visible. Streamlines of the z-plane (below, left) show two vortices. A large one at $ x / h \approx 0.4 $, $ y / h \approx -0.4 $ and a smaller one at $ x / h \approx 0.7 $, $ y / h \approx 0.5 $. These vortex positions once more reflect the asymmetry. The isosurface of the total pressure also shows a slight skewness in this plane. 

$ \varphi = 90 \degree $: The upper part of the vortex ring has increased in height ($z$-direction) in the y-plane. The entire wake is more symmetric in all views and the streamlines behind the dead water are hardly deflected in comparison to the previous and other phases. Thus the saddle point moves downwards ($ z / h \approx 0 $) and closer to the base ($ x / h \approx 1.2 $). The vortices in the x-plane have changed to a symmetric arrangement. Their focuses are located at middle height ($ z / h \approx 0 $, $ y / h \approx +/-0.25 $).

$ \varphi = 180 \degree $: The streamlines show that the upper part of the vortex ring in the y-plane becomes very small and shifts closer to the base again compared to other phases. Based on the streamlines in this plane the saddle point is located at $ x / h \approx 1.2 $, $z / h > 0.4$ and hence moved upwards. The isosurface of the total pressure essentially follows the limitation of the dead water by the streamlines except in the upper section of the vortex ring. An extension of the total pressure isosurface encloses the vortex in the upper section of the dead water. In the $x$-plane an asymmetric situation appears again. As in $ \varphi = 0 \degree $ the total pressure in this plane is distributed asymmetrically. On inspection of the streamlines in the dead water in the $z$-plane this asymmetry is also visible. The vortices in the x-plane have changed to an asymmetric arrangement. Their focuses are located at $ z / h \approx 0.1 $, $ y / h \approx -0.45 $ and $ z / h \approx 0.3 $, $ y / h \approx 0.15 $. In contrast to phase $ \varphi = 0 \degree $, the upper vortex is now on the right side. 

\clearpage

\clearpage

\clearpage

$ \varphi = 270 \degree $: In the y-plane the saddle point has moved downwards and closer to the base ($ x / h \approx 1.0 $, $ z / h \approx 0.3 $). In the section behind the dead water the streamlines are deflected far upwards from the underbody region in order to be guided back into the horizontal with an S-stroke near the saddle point. At this point, strongly curved, but not completely closed streamlines appear. These curvatures of the streamlines behind the saddle point are an evidence for a vortex at $ x / h \approx 1.25 $. A closer inspection of the upper part of the dead water of this plane reveals another small vortex at $ x / h \approx 0.5 $. In comparison to phase $ \varphi = 180 \degree $ the size of this vortex is increased. The location of the saddle point indicates that the downstream vortex part has separated from the upstream part and left the boundary of the dead water. A clue to a vortex shedding from the dead water. The streamlines and the isosurface of the total pressure in the x-plane are more symmetric than phases $ \varphi = 0 \degree $ and $ \varphi = 180 \degree $. However, at the $x$-plane (upside, right side) streamlines show two vortices asymmetric in size and position. The isosurface of total pressure in the $x$-plane shows that the downstream section of the upper vortex is concentrated in the middle (in $y$-direction) area ($-0.25 < y / h <0.25$, $0.2 < z / h <0.4$). This is also noticeable in the z-plane. In this view, the upper part of the isosurface forms a small triangular branch. This indicates that the effect of the downstream vortex in the upper shear layer is limited to the range of about $-0.25 < y / h <0.25$. Similar observations can be seen by looking at the different $y$-planes of the measurement (not shown here, \citep[see][]{BockDiss}).

In the $y$-plane the limiting surface of the total pressure correlates well with the limitation of the dead water formed by the streamlines of the upper and lower sides in the $y / h = 0$ plane. Together with the upper vortex the dead water is squeezed relative to the base. However, elongated extensions of the isosurface of the total pressure are formed. These start from $ x / h \approx 1 $ and $ z / h \approx 0.1 $ and extend to $ x / h \approx 1.4 $ and $ z / h \approx 0.4 $. These extensions can also be observed in the $x$-plane. There they are located at external positions $ y / h \approx +/- 0.4 $. The streamlines in this plane show that they are surrounded by two vortices. In the $x$-plane the flow is symmetrical. In the $z$-plane the flow field (both isosurface and streamlines) is symmetrical as well. The extensions of the isosurface of the total pressure are also enclosed by vortices in the $z$-plane. They thus form a horseshoe vortex which starts from the vortex ring in the dead water and closes behind the dead water.

Especially the streamlines in the $y$-plane allow the comparison with the phase states in Figure \ref{fig:RekoEbene} and therefore a comparison of the reconstruction of simulation and experiment. As the phase states $ \varphi $ of experiment and simulation may not be perfectly synchronized, an exact match of the streamlines can not be expected. The phases considered in the simulation may lie somewhere between the phases shown in the experiment. Hence, they may represent transition states between the phases in experiments. However, the streamlines of the $y$-plane from the experiments show that the main variations are caused by the changes of the size of the upper part of the vortex ring in the dead water. The experimental results also show a related motion of the saddle point. In accordance with the simulation the saddle point moves up-/downwards and away/towards the base during the process. In all observations this is accompanied by a formation of strong curvature of streamlines behind the saddle point. These curvatures indicate a vortex detached from the dead water. In conclusion a qualitatively good correspondence of the streamlines in the $y$-plane between experiment and simulation (cf. Figure \ref{fig:RekoEbene} below left) is given. 

\clearpage

\subsection{Low frequency wake modes}

%\clearpage

\begin{figure}[h]
  \centering
      \begin{subfigure}[]{\textwidth}
          \centering
          %\rule[1ex]{\textwidth}{0.4pt}
          \caption*{maximal amplitude}
          \includegraphics[width=12cm]{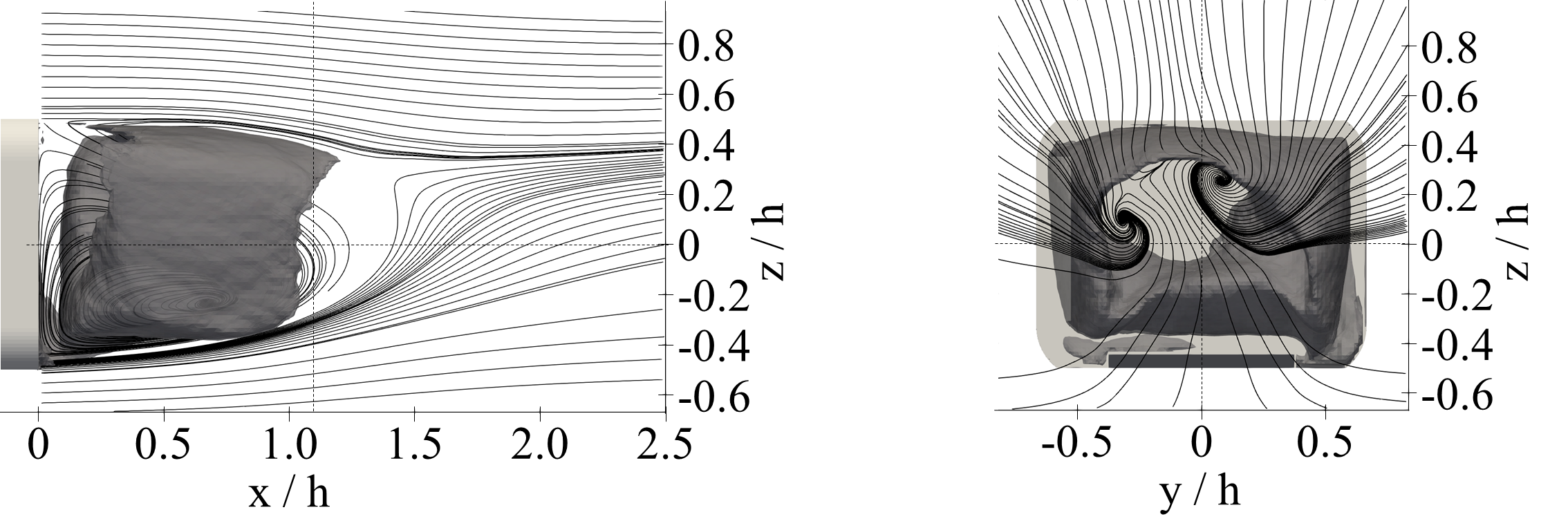}
          \rule[1ex]{\textwidth}{0.4pt}
      \end{subfigure}
      \begin{subfigure}[]{\textwidth}
          \centering
          \caption*{minimal amplitude}
          \includegraphics[width=12cm]{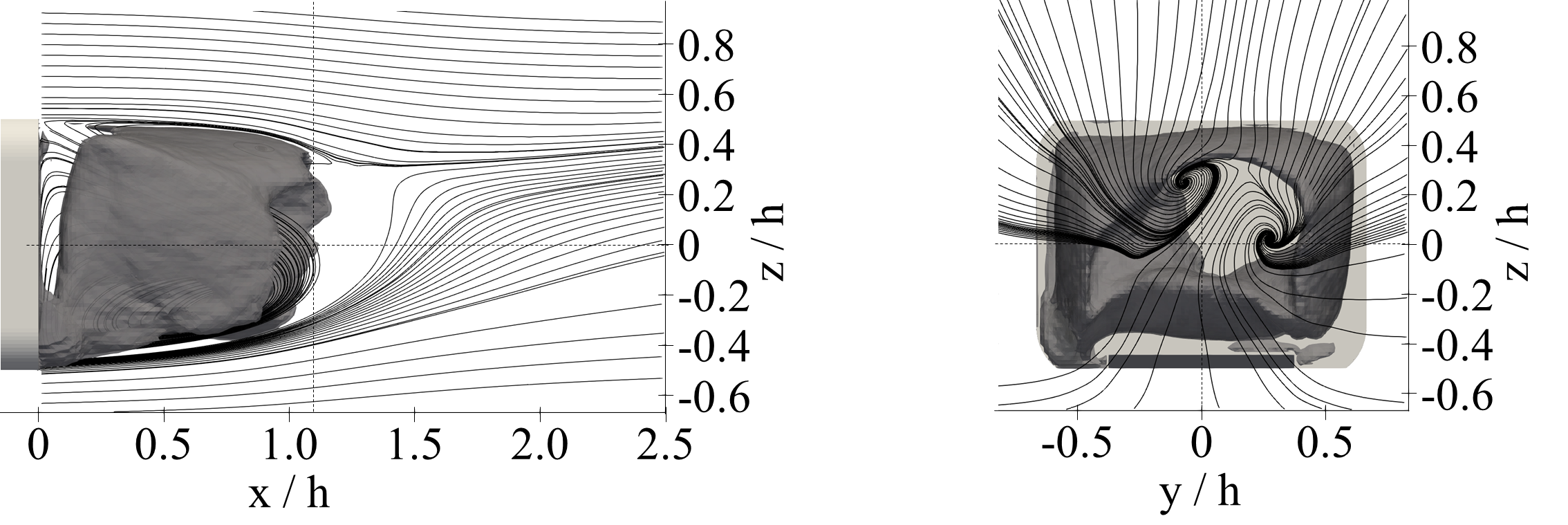}
          \rule[1ex]{\textwidth}{0.4pt}
      \end{subfigure}
  \caption{LOM by reconstruction of POD mode $m=1$ of the wake volume sampled with $ f_s = 30 Hz $ and in superposition with the averaged flow. Isosurfaces of $ c_{p,t} = -0.1 $ from simulation.}  
\label{fig:RekoVolLow}
\end{figure}

The low-frequency fluctuations of the base pressure distributions on the SAE model represent a flapping motion in width direction. To consider the form of the low-frequency motion, the wake volume reconstruction of mode $m=1$ from the POD decomposition of the long-term sampling at $ f_s = 30 \hspace{5pt} Hz $ ($ Sr_h = 0.18 $) and moving average from the simulation is used. The influence of fluctuations from the high and medium frequency ranges is eliminated by the moving average. The temporal amplitudes of this mode fluctuate broadband in the range of the lowest resolved frequencies $ Sr_h < 0.04 $ as shown in \citet[][]{BockDiss}. Figure \ref{fig:RekoVolLow} shows the reconstruction of the first mode in the wake volume using the maximum and minimum temporal amplitudes or a positive or negative deflection around the time-averaged flow field. In the picture, divided into two rows, the positive and negative deflections are shown with $2$ different views each. 
The reconstruction at maximum deflection of Mode $m=1$ looks very similar to the averaged flow (cf. Figure \ref{WakeAv}) when viewed from the side. The upper part of the vortex ring is significantly smaller than the lower part and at the same time has a larger distance to the base. Thus the saddle point is at $ z / h \approx 0.4 $ and the streamlines from the underside are strongly deflected upwards before they are deflected horizontally at this point. The isosurface of the total pressure encloses the vortices and thus essentially the dead water. In the view from the rear ($x$-plane) a strong asymmetric flow field is shown. The two axial vortices in the considered plane are shifted to the left ($y / h < 0$). Additionally, the curvature of the total pressure in the upper half is also shifted in this direction. In the side plane ($y$-plane) of the reconstruction of the minimum deflection of mode $m=1$, the flow field changes only insignificantly. In this case, the view of the $x$-plane corresponds to a mirroring around the $y$-plane.

%\clearpage

%
%
\section{Discussion}
The data from the investigations in the wake of the SAE model show coherent structures of different time scales. The traces of the KH vortices in the velocity fluctuations of the shear layer can be detected at a relatively high frequency. In addition, the vortex shedding in the medium frequency range can be traced back not only in the wake volume but also in the velocity fluctuations in the shear layer down to the base pressures. Moreover, a pronounced deflection of the wake can be recorded in all measurement data, which takes place in the very low frequency range. The interrelationships of the fluctuation motion of the vortex shedding and the deflection of the wake in the width direction are discussed in more detail in the following.

\subsection{Vortex shedding}
The vortex shedding on the SAE model can be interpreted as a meander-shaped process in the upper half of the wake. The changes in the flow fields in Figure \ref{fig:RekoVol1} and Figure \ref{fig:RekoVol2}, which take place in time scales of $ Sr_h \approx 0.22 $, can be interpreted as continuous process, where phase $ \varphi = 0 \degree $ is followed by phases $ \varphi = 90 \degree $, $ \varphi = 180 \degree $, $ \varphi = 270 \degree $, $ \varphi = 0 \degree $, and so on. 

Starting from $ \varphi = 270 \degree $ to $ \varphi = 0 \degree $, the size of the upper part of the vortex ring becomes smaller in the streamlines of the $y / h = 0$ plane. The upper part of the vortex ring becomes larger in phase $ \varphi = 90 \degree $ both in the streamlines of the $y / h = 0$ plane and in the isosurface of the total pressure. In the following phase $ \varphi = 180 \degree $ it is again of reduced size. Hence, the vortex detachment starts between $ \varphi = 180 \degree $ and $ \varphi = 270 \degree $. 
The vortex shedding is connected with a motion of the barycenter of the total pressure in the upper domain from right ($y / h > 0$) to left ($y / h < 0$). This is concluded by observation of the total pressure isosurface and the streamlines in the $x$-planes. In phases $ \varphi = 90 \degree$ and $ \varphi = 270 \degree$ the flow field in this plane is symmetric. In the other two phases the total pressure and the streamlines are asymmetrically distributed. From visual observations of the total pressure isosurface of phase $ \varphi = 0 \degree$ a concentration of the upper part of this surface in the right section ($y / h > 0$) is concluded. In phase $ \varphi = 180 \degree$ the total pressure isosurface is concentrated in the opposing section ($y / h < 0$). This observation of lateral motion of the dead water in the $y \approx 0$ plane may cause an increase and decrease of the upper part of the vortex ring during the observed process. Nevertheless, the curvatures of streamlines behind the dead water in the $y$-plane mark the shedding of a vortex during the same process. In phase $ \varphi = 270 \degree$ branches of the isosurface of the total pressure at medium height ($z / h = 0.1$) surrounded by vortices (streamlines in the $x$- and $z$-planes) indicate a horseshoe-shaped vortex. At a further state of the process ($ \varphi = 270 \degree $) this area within the total pressure isosurface is shown as a continuous area and branches move upwards ($ z / h \approx 0.25 $) and become narrower (in $y$-direction). The switch of the concentration of the total pressure isosurface during the vortex shedding suggests a meandering detachment of the vortex of the upper part of the vortex ring.

This describes a vortex shedding process in the wake. It is obvious that this is a vortex which maintains a connection through a horseshoe vortex during the shedding process from the vortex ring (at least for a certain period of time). Such a behavior makes sense as Helmholtz's law of circulation remains fulfilled. With increasing distance from the base after leaving the dead water, the detached vortex structure is more difficult to detect. This suggests the assumption that this will quickly vanish due to diffusion in the far wake. As a result, the vortex would have to expand and lose intensity after leaving the dead water.

\begin{figure}[h]
  \centering
          \includegraphics[width=12cm]{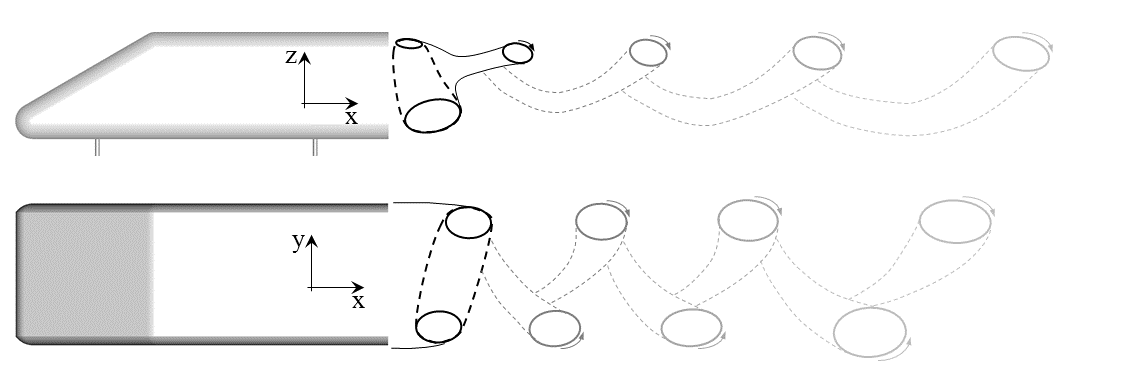}
  % Images in 100% size
  \caption{Schematic illustration of the vortex shedding at the SAE squareback vehicle.}
\label{fig:VortexShedding}
\end{figure}

Vortex shedding on the SAE model can be considered as the motion of a vortex loop in the upper half of the wake. Figure \ref{fig:VortexShedding} schematically shows the proposed vortex shedding in the wake of the SAE model. The process is outlined in two views (top: $y$-plane and bottom: $z$-plane). The border of the dead water is shown as a black, solid line and the vortex ring is shown as black circles. Dashed lines represent borders of the vortices that do not lie in the plane of observation. At the $y$-plane a vortex starts to separate from the upper section of the dead water. Previously detached vortices can be seen  further downstream. These vortices are getting bigger, weaker and increase their distance to each other the further they move downstream from the base. The increase in the distance of the vortices current distance is concluded from the different frequencies of the vortices \citep[cf.][]{Barros2015, barros2017effects} of the vortex shedding depending on their orientation (interaction of the shear layers with the distance $h$ or $b$). The increasing greying out of vortices embodies the decrease in the strength of the vortices. All detached structures are still connected to each other. By view from above, this connection represents two vortices shifted in flow direction at opposite sides of the vehicle. These vortices within this plane always lie between two vortices in view from the side. The offset of these vortices describes the observed flapping during vortex shedding.

The vortex shedding frequency is mainly measurable in the shear layer and is noticeable via the flow field up to the base pressures at $ Sr_h \approx 0.2 $. The strongest fluctuations in the shear layer mainly occur with $ Sr_h \approx 0.2 $ in the vicinity of the saddle point. This is where the greatest fluctuations in vortex shedding occur. The reconstruction of the flow volume is based on POD modes (from $m=1$ to $m=4$) whose temporal amplitudes have a peak in the frequency domain $ Sr_h \approx 0.22 $. The measured frequencies in the flow field seem somewhat higher at first glance. However, these frequencies originate from very broadband peaks.

The effects of vortex shedding can be traced back to the motions of the barycenter of the base pressure and in the flow volume, but above all in the shear layer. The spectra of the dimensionless, spatially averaged surface pressure gradient show frequency peaks similar to those in the shear layer and in the POD modes of the flow field (cf. Figure \ref{SpektrenBasis}). Thus, the dominant frequencies in the shear layer in the range $ Sr_h \approx 0.2 $ and the POD modes of vortex shedding also occur in the observations of pressure fluctuations.

Due to the position and the size of the vortex, the base pressures only slightly reflect the shedding of the vortex. The frequency $ Sr_h \approx 0.2 $, which is assigned to vortex shedding, is broadband and not dominant in the spectra of the barycenter of motion of the base pressure distribution. Broadband capability can be considered typical for turbulent flows. The weak development of the peaks can be explained by observations of vortex shedding in the flow field.

This weak response is due to relatively small vortices far away from the base, which is intensified by the diffuser. Vortices act as pressure sinks on surfaces. The closer a vortex is to the surface or the larger the vortex, the lower the pressure on the surface. The base pressure distributions show the lowest pressures in the lower section. This correlates with the observation from the temporally averaged flow field, where in this area the vortex ring is very pronounced and close to the base. However, the vortices of vortex shedding are comparably small and far away from the base. This is because vortex separation mainly takes place in the upper dead water and close to the saddle point. Thus these structures are located far from the base surface compared to the position of the lower part of the vortex ring or the length of the dead water area. In addition, the structures observed are smaller than half the height of the base surface. The coherent structures of other flows, such as the K\'arm\'an Vortex Street in a two-dimensional flow, have a much stronger effect on the base. In the wake of the SAE model, the size and strength of the vortices is much smaller and the distance to the base is larger. It is obvious that these properties of the size of the detached vortices and the distance to the base is even further increased with the diffuser by the wake topology.

With other three-dimensional bluff body flows and squareback vehicle models, the vortex separation is also meander-shaped. However, the meander-shaped detachment of the vortices is not limited to a part of the dead water, as is assumed in the SAE model due to the influence of the diffuser. A meander-shaped vortex separation was proposed and documented for many other three-dimensional bluff body flows, such as the circular or elliptical disc  \citep[][]{Kiya1999turbulent, yang2015low}, the sphere  \citep[][]{Berger1990} and squareback vehicle models  \citep[][]{Grandemange2013, duell1999experimental}. The frequencies of the vortex shedding are all in the range $Sr = 0.14...0.22$, in which the vortex shedding frequency is also for the SAE squareback model. Compared to the vortex shedding of the circular or elliptical disc \citep[][]{Kiya1999turbulent, yang2015low} and to the documented squareback vehicle models, without diffuser \citep[][]{Grandemange2013, duell1999experimental}, the affected area is significantly smaller in relation to the cross sectional area. In addition, the coherent structures of vortex shedding move alternately in all directions in the wake of the circular disc \citep[][]{yang2015low} and of the squareback vehicle model \citep[][]{duell1999experimental}. In contrast, the coherent structures of vortex shedding on the SAE model only move away from the ground in the vertical axis. As a special feature of the flow around an SAE squareback model and unlike the behavior of the Ahmed squareback model \citep[][]{Grandemange2013} it is shown that almost exclusively the upper vortex is involved in the vortex shedding. The area of the lower part of the vortex ring remains relatively stable. It is obvious that the influence of the diffuser on the velocity profiles in the wake is responsible for the stabilization of the lower area of the wake.

\subsection{More realistic geometries and vortex shedding}
\begin{figure}[h]
  \centering
  \includegraphics[width=12cm]{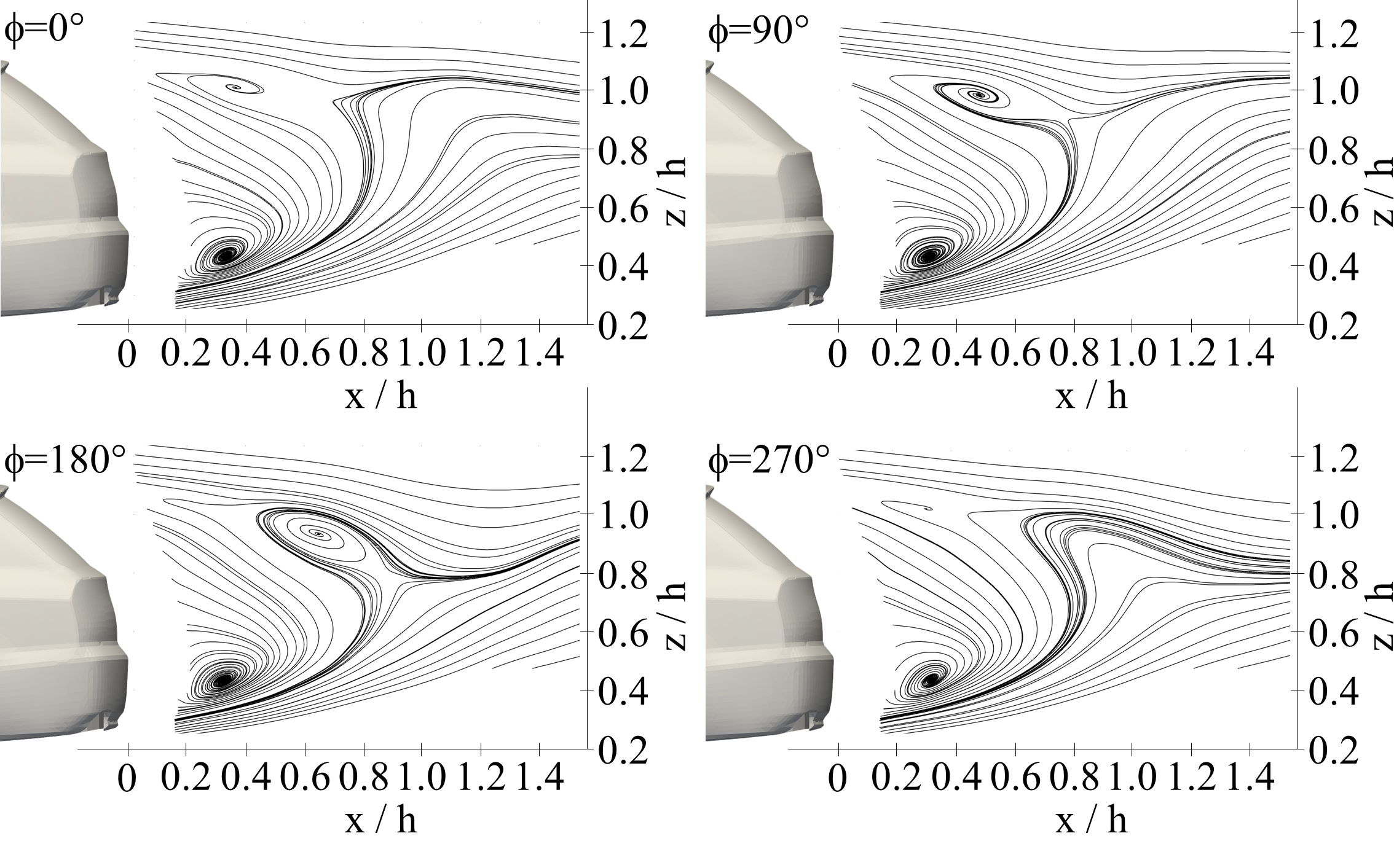}
  % Images in 100% size
  \caption{POD reconstruction of the wake flow on a DrivAer model in the $y / h = 0$ plane.}
\label{fig:RekoDrivAer}
\end{figure}

The meander-shaped vortex shedding in the upper half of the wake can be regarded as a basic form in connection with the vortex shedding on many squareback vehicle types. The geometry of the SAE model contains only the most outstanding features of a squareback vehicle. The behavior of the meander-shaped vortex shedding on the SAE model, which only takes place in the upper half of the wake, can also be shown on the more complex DrivAer model geometry. For this purpose, the $y / h = 0$ plane in the wake of the $1$:$4$ DrivAer model with a Reynolds number of $ Re_l = 3.6 \hspace{5pt} 10^6 $, was measured with PIV as on the SAE model.  The basic vehicle geometry is described by \citet[][]{heft2012introduction} and the configuration of these measurements and some details about the cooling air flow geometry by \citet[][]{Kuthada2016effects}. The flow fields were decomposed with the POD. The first two modes also represent a convective pair of modes. Figure \ref{fig:RekoDrivAer} shows the reconstruction of this mode pair in four phase positions with streamlines. The representation is analogous to the reconstruction of the flow field in the plane on the SAE model (Figure \ref{fig:RekoEbene}). The rear end of the vehicle model can be seen on the left in each subframe. It is noticeable that the saddle point is closer to the base or not as far from the rear edge as with the SAE model. This is a feature that can be attributed to the continuous shape optimization of the production vehicles that is taken into account in the DrivAer model. However, the wake flow characteristics and the wake flow of the SAE model both have a greater curvature of the lower free streamline compared to the upper one. This provides a stronger upwards ambition for this streamline. Thus the basic shape and the arrangement of the vortex system in the dead water is very similar to the SAE squareback model, although the base of the DrivAer model has very different degrees of surface curvature gradients. Streamlines thus show the lower vortex in almost same position and unchanged size at all four phase positions. The upper, smaller vortex shows an increase and a downstream convection up to a detachment as with the SAE model. The vortex shedding of the DrivAer is therefore very similar to the SAE model. Together with the fact that in all documented cases of three-dimensional bluff body flows a meander-shaped vortex shedding occurs, it is very likely that in the DrivAer model the vortex shedding is also meander-shaped in the upper half of the dead water. This leads to the hypothesis that this vortex separation is basically valid for the majority of squareback geometries and possibly also Sport Utility Vehicles (SUV).

\subsection{Low frequencies}
The low-frequency fluctuations of the wake represent a flapping that may interact with the vortex shedding. For the flow around three-dimensional bluff bodies, such as the circular or elliptical disc \citep[][]{Kiya1999turbulent, yang2015low}, the sphere \citep[][]{Berger1990} and squareback vehicle models \citep[][]{Grandemange2013, duell1999experimental}, a low frequency range of the following motion is documented parallel to the vortex shedding. The spectra of the dimensionless, spatially averaged surface pressure gradient and the flow field modes on the SAE model have very pronounced and dominant fluctuations in the low frequency range. These are also visible in the spectra of the flow field of the shear layer. In the spectra of the dimensionless, spatially averaged surface pressure gradient (Figure \ref{SpektrenBasis}) two ranges of accumulations of fluctuations can be distinguished. These are the ranges $ Sr_h \approx 0.07 $ and $ Sr_h < 0.06 $. The latter can only be seen in the dimensionless, spatially averaged surface pressure gradient in $y$, i.e. in the motions of the barycenter of pressure in $y$-direction. This is to be interpreted as a deflection of the dead water mainly in the $y$-direction (i.e. the width), which takes place in very long time scales. When comparing the POD reconstruction of the flow field in the mid-frequency range (cf. Figure \ref{fig:RekoVol1} and Figure \ref{fig:RekoVol2}), no vortex shedding can be seen when looking at the low frequencies. Rather, it is a lateral motion of the wake, combined with compression or stretching of opposite areas in the width (y-direction) and can therefore be described as flapping. According to \citet[][]{brackston2016stochastic}, the fluctuations at low frequencies or long time scales correspond to the range to be assigned to the motions of the quasi-random fluctuations and bi-stabilities. The distributions of the barycenter of pressure positions (cf. Figure \ref{ScatterSP}) show that this is not a bi-stability. From the perspective of the spatial motion pattern, this flapping motion is comparable to the bi-stability of the Ahmed model, but with different temporal behavior. The time scales of this motion are similar to flapping, as with the circular disc \citep[][]{yang2015low} or the axial symmetric blunt body \citep[][]{Gentile2016afterbody}. Similar to \citet[][]{yang2015low} describing the modulation of vortex shedding by flapping, the two corresponding phenomena could interact with each other on the SAE model.

The absence of bi-stability is in accordance with the findings of \citet[][]{Grandemange2013a}. 
Here, different combinations of aspect ratio and ground clearance of an Ahmed model were examined with regard to bi-stability. Measured against these parameters, the SAE squareback model is just within the range of an occurrence of a bi-stability. However, it should be noticed that the wake flow of the SAE model, especially through the diffuser, differs from that of the Ahmed model. The diffuser causes the wake to move upwards. The flow topology is therefore rather comparable to that of the Ahmed model with an aspect ratio $h / b = 0.75$ and a ground clearance $c / h = 0.1666$ \citep[compare][]{Grandemange2013a}. This corresponds to a case without occurring bi-stability.

\subsection{Interaction of low frequencies vs vortex shedding}
The detachment of a horseshoe vortex in the upper shear layer correlates well with the distribution of the base pressure fluctuations which have their maximum in the middle of the upper half. Therefore it is surprising at first that the shedding of the vortex is not better visible in changes of the base pressure distributions. One possible cause is that the velocity fluctuations of vortex shedding are poorly transferred to the base. Reasons could be the size, strength and distance of the vortex compared to the base or the spatial phase shift of the vortex. Another cause could also be the interaction of vortex shedding with the low-frequency motions.

Some of the current research shows two processes that influence each other or may even be linked to each other \citep[][]{rigas2017weakly, barros2017effects}. In many flows of blunt bodies, low-frequency behavior was found \citep[][]{Grandemange2013, brackston2016stochastic, duell1999experimental, Gentile2016afterbody, Rigas2014, Kiya1999turbulent, yang2015low, Grandemange2012, Grandemange2012a}. The low-frequency flapping motion observed in this work resembles the low-frequency flapping of axial symmetric blunt bodies or considering the amplitudes the bi-stability of some squareback vehicle models \citep[][]{Grandemange2013, Grandemange2013a}.

Many indications show that the vortex shedding interacts at least with the low-frequency behavior. In some works \citep[][]{Kiya1999turbulent, yang2015low} it is shown that the low-frequency motion in the wake of axial symmetric blunt bodies causes a modulation of the vortex shedding. Furthermore, the work of \citet[][]{rigas2017weakly} on the axial symmetric blunt body and of  \citet[][]{Grandemange2013} on the Ahmed model show that the vortex shedding is aligned with the low-frequency wake deflection.Another observation underlines that the interaction of the coherent structures of vortex shedding and low-frequency motions cannot be fully observed separately. \citet[][]{Rigas2014} investigated the coherent structures on the axial symmetric blunt body by the barycenter of pressure motion, POD and Fourier decomposition of the base pressure distribution, whereby the vortex shedding could not be separated from the low-frequency motions in all methods. The findings of \citet[][]{Gentile2016afterbody, Gentile2016low}, the POD and the barycenter of pressure motion in wake planes lead to the same results in this respect.

The findings in this work and the comparison with documented research show that there is the possibility that the low-frequency modulation frequency may result from a dislocation during vortex shedding as in the two-dimensional flow \citep[][]{williamson1992natural}. The motion of the barycenter of pressure on the SAE model observed here is reminiscent of the constellation in the two-dimensional flow. In the two-dimensional flow, the barycenter of pressure moves in the direction perpendicular to the two-dimensional flow through the low-frequency dislocation. At the vertical axis of two-dimensional flow, the barycenter of pressure will move only by vortex-shedding \citep[][]{BockDiss}. Likewise, the low-frequency motion of the barycenter of pressure on the SAE model mainly takes place in a direction perpendicular to the main direction of a motion of the barycenter of pressure as a result of the vortex shedding. Another analogy is that in the two-dimensional flow the low-frequency fluctuation in the base pressures is more easily detected than the vortex shedding. The low-frequency fluctuations are recorded at each measuring point. However, the fluctuations based on vortex shedding only become clearer by looking at certain measuring points or through certain filters \citep[][]{BockDiss}. Finally, the results show another characteristic that indicates a strong interaction or even combination of both processes. Since the vortex shedding on the SAE model is phase-shifted over the width, it is accompanied by a flapping motion. It is very likely that an additional (low-frequency) flapping motion changes this phase shift and thus also acts as a dislocation. Nevertheless, the dislocation seems unlikely or has not yet been determined as the cause for modulation on the basis of various studies \citep[][]{Kiya1999turbulent, yang2015low}.

On the basis of this work and other publications, a clear statement cannot be made on whether dislocation does or does not occur. A modulation of vortex shedding is very obvious as many authors conclude from their results. Dislocation, on the other hand, cannot be clearly demonstrated. Further investigations would be necessary to prove this. For example, structures of vortex shedding during shedding could be traced to detect dislocations if necessary.
\section{Conclusion}
In the present study, the coherent structures in the wake of a squareback SAE model were examined experimentally and numerically. Time-resolved base pressures  and velocities in different planes were measured statistically with pressure transducers and PIV. Using VLES simulations, additional time resolved and spatially coherent data were generated in the wake of the SAE model. Small-scale fluctuations were spatially filtered by observing barycenters of pressure, dimensionless, spatially averaged surface pressure gradients at the base and in the flow field by POD. An additional differentiation of time scales was achieved by spectra.

It has been shown that the shape of the vortex ring in the wake can differ on the different squareback vehicle models. These differences are probably geometry-dependent. This has an effect on the vortex shedding, on base pressure fluctuations and on the behavior of fluctuations in the low frequency range. It can be assumed that these flow properties must be taken into account in flow control measures that affect the behavior of the fluctuations. In addition, the potential to reduce aerodynamic drag could also change due to the different behavior of the flow situation.

The structures of vortex shedding are difficult to detect, especially in the base pressures of the SAE squareback model, which can nevertheless be used as feedback sensors by means of suitable filtering. The dead water of the SAE squareback model consists of a vortex ring which is located close to the base in the lower part and is deformed by the diffuser in the upper part. The shape of the vortex ring and thus the dead water area differs from that of an Ahmed squareback model. The vortex shedding on the SAE model can be interpreted as a meander-shaped process in the upper half of the wake with an average frequency of $ Sr_h \approx 0.2 $. Due to the position and the size of the vortex, the base pressures only slightly reflect the shedding of the vortex. This weak response is due to relatively small vortices far away from the base. As a feedback sensor for active flow control of vortex shedding, the base pressures are therefore not optimal. Filtering the sensor signals via the barycenter of pressure can improve this problem. In the context of the investigations of \citet[][]{brackston2016stochastic} it can be seen that the signal of the barycenter of pressure is suitable to influence the vortex shedding at the Ahmed model in a feedback loop, even though it was excited and not reduced by this example and thus probably the aerodynamic drag was increased.

The meander-shaped vortex shedding in the upper half of the wake can be regarded as a basic form in connection with the vortex shedding on squareback vehicle types. This is confirmed by the meander-shaped vortex shedding in other three-dimensional blunt bodies and squareback vehicle models \citep[][]{Grandemange2013, barros2017effects}. However, the meander-shaped shedding of the vortices is in these cases, in contrast to the observed vortex shedding of the SAE squareback model not limited to a part of the dead water. This distinction is attributed to the influence of the diffuser. 

The low-frequency motions of the wake represent a flapping in width direction. It is suspected that the flapping may interact with the vortex shedding. The lack of bi-stability is in accordance with the findings of \citet[][]{Grandemange2013a}. Separation of low frequency, or detection of dislocations remains to be investigated in further work.

The exact form of vortex sheddings could be relevant for the positions of an active flow control measure. The findings of this study show that there are strong influences of geometry on the fluctuation motion in the low frequency range, but also on the vortex shedding. The understanding of the vortex shedding process on the SAE model and on squareback vehicle types suggests the response to active control. The shedding of the vortex loop and the reaction to a control approach should be basically very similar in the axial symmetric body \citep[][]{rigas2017weakly} and the Ahmed model \citep[][]{barros2017effects}. \citet[][]{barros2017effects} showed a correlation of the sensitivity of vortex shedding to the flow control position with the preferred direction of vortex shedding. In the vortex shedding of the SAE model, the activity range is limited to the upper half of the wake compared to these flows. For such an application, only a reduced range of pulsed or synthetic jets would have to be excited (possibly only on the side and the top edges). At the same frequencies and amplitudes, similar results can be expected as for \citet[][]{barros2017effects} and \citet[][]{rigas2017weakly}. Up to now, however, in doing so aerodynamic drag has been increased and the energy content of the fluctuating motions of the low frequencies reduced. Nevertheless, the findings of \citet[][]{barros2017effects} show that a relevant reduction of the aerodynamic drag forces is possible by control approaches  on the vortex shedding.

Another consequence of the results of this study is that the form of vortex shedding could change due to passive or active control approaches of the flow field. The findings of vortex shedding on the SAE model compared to the Ahmed model show the possible influence of geometry on coherent structures. This also means that an influence on the temporally averaged flow in the wake, e.g. by any means of aeroshaping, could severly change the process of vortex shedding again.

Further detailed studies on the coherent structures of squareback vehicle models or three-dimensional blunt bodies with different geometries could help to clarify the following questions regarding the impact on aerodynamic forces. What is the effect of the fluctuations and how high are the energy components of the coherent structures? In applications this would be relevant for the influence on aerodynamic drag, but also on other forces, e.g. in driving dynamics. This could point the way to show how high the energy input of an active control measure could be, so that the energy saved is still worthwhile in optimising the behavior of a coherent structure. The present study suggests that this is to be expected with significantly lower energy proportions compared to the two-dimensional flow during vortex shedding.

A flow control measure, such as the place of excitation in active flow control, must depend on where the processes take place and which is the main direction of motion. The present investigations have shown that this can, however, depend on the geometry. A more systematic investigation of geometric influences such as different diffuser angles, rear shape and possibly others could provide further important information here.

In conclusion, these results improve the understanding of the behavior of coherent structures and the effect on base pressure, i.e. vehicle forces for squareback vehicle types. This will be crucial in developing new strategies to influence and control vehicle forces.

The author acknowledges the support of the IVK at the University of Stuttgart where the experiments and simulations were performed. Furthermore, the support of the Friedrich-und-Elisabeth-BOYSEN-Stiftung is thankfully acknowledged for the funding under grant BOY11-No.77. The author also warmly thanks Jochen Wiedemann, Nils Widdecke, Timo Kuthada, Christoph Sch\"onleber, Alexander Hennig, Max Tanneberger and Daniel Stoll for fruitful discussions.

\bibliographystyle{plainnat}
\bibliography{Benjamin_Bock_CoherentStructuresInTheWakeOfASAESquarebackVehicleModel}

\end{document}